\documentclass[11pt]{article}

\newsavebox{\foobox}
\newcommand{\slantbox}[2][0]{\mbox{%
        \sbox{\foobox}{#2}%
        \hskip\wd\foobox
        \pdfsave
        \pdfsetmatrix{1 0 #1 1}%
        \llap{\usebox{\foobox}}%
        \pdfrestore
}}
\newcommand\unslant[2][-.25]{\slantbox[#1]{$#2$}}

\newcommand{\mpi}{\text{\unslant[-.18]\pi}}
\newcommand{\mdelta}{\text{\unslant[-.18]\delta}}

\usepackage[left=2cm, right=2cm, top=2.5cm, bottom=2.5cm]{geometry}
\usepackage[x11names]{xcolor}
\usepackage{fancyhdr, amssymb, cancel, amsmath, graphicx, pgfplots, tikz}
\usepackage{isomath}

\usetikzlibrary{shadows}

\newcommand{\stylecolor}{blue!50!black}

\usepackage[labelfont={bf,sf, color=\stylecolor}, margin={1.5cm,0cm}]{caption}

\usepackage[colorlinks=true, urlcolor=\stylecolor!70!white, linkcolor=\stylecolor, citecolor=\stylecolor!70!white, hyperindex=true, linktocpage=true]{hyperref}

\usepackage[explicit]{titlesec}

\newcommand*\sectionlabel{}
\titleformat{\section}
  {\gdef\sectionlabel{}
   \Large\bfseries\scshape}
  {\gdef\sectionlabel{\thesection }}{0pt}
  {\begin{tikzpicture}[remember picture,overlay]
       \end{tikzpicture}
  }
\titlespacing*{\section}{0pt}{0pt}{0pt}

\newcommand*\subsectionlabel{}
\titleformat{\subsection}
  {\gdef\subsectionlabel{}
   \large\bfseries\scshape}
  {\gdef\subsectionlabel{\thesubsection  }}{0pt}
  {\begin{tikzpicture}[remember picture,overlay]
    	\draw (-0.15, 0.02) node[right] {\color{\stylecolor} \textsf{\subsectionlabel}};
	\draw (1.25, 0) node[right] {\color{\stylecolor} \textsf{#1}};
	\fill[color=\stylecolor] (1,-0.25) rectangle (1.1, 0.25);
       \end{tikzpicture}
  }
\titlespacing*{\subsection}{0pt}{10pt}{10pt}

\newcommand*\subsubsectionlabel{}
\titleformat{\subsubsection}
  {\gdef\subsubsectionlabel{}
   \bfseries\scshape}
  {\gdef\subsubsectionlabel{\thesubsubsection.\ \  }}{0pt}
  {\begin{tikzpicture}[remember picture,overlay]
    	\draw (-0.15, 0) node[right] {\color{\stylecolor} \textsf{\subsubsectionlabel#1}};
       \end{tikzpicture}
  }
\titlespacing*{\subsubsection}{0pt}{7pt}{7pt}

\pgfplotsset{every axis legend/.append style={at={(.5,0.95)},anchor=north west}}

\begin{document}

\allowdisplaybreaks

\pagestyle{fancy}
\renewcommand{\headrulewidth}{0pt}
\fancyhead{}

\fancyfoot{}
\fancyfoot[C] {\textsf{\textbf{\thepage}}}

\begin{equation*}
\begin{tikzpicture}
\draw (\textwidth, 0) node[text width = \textwidth, right] {\color{white} easter egg};
\end{tikzpicture}
\end{equation*}

\begin{equation*}
\begin{tikzpicture}
\draw (0.5\textwidth, -3) node[text width = \textwidth] {\huge  \textsf{\textbf{Hydrodynamic charge and heat transport % in quantum critical fluids
      on \\ \vspace{0.07in}  inhomogeneous curved spaces}} };
\end{tikzpicture}
\end{equation*}
\begin{equation*}
\begin{tikzpicture}
\draw (0.5\textwidth, 0.1) node[text width=\textwidth] {\large \color{black} \textsf{Vincenzo Scopelliti,}$^{\color{\stylecolor} \mathsf{a}}$ \textsf{Koenraad Schalm,}$^{\color{\stylecolor} \mathsf{a}}$ \textsf{and Andrew Lucas}$^{\color{\stylecolor} \mathsf{b}}$};
\draw (0.5\textwidth, -0.5) node[text width=\textwidth] { $^{\color{\stylecolor} \mathsf{a}}$  \small \textsf{Instituut-Lorentz for Theoretical Physics, Leiden University,
Niels Bohrweg 2, Leiden 2333CA, The Netherlands}};
\draw (0.5\textwidth, -1) node[text width=\textwidth] { $^{\color{\stylecolor} \mathsf{b}}$  \small\textsf{Department of Physics, Stanford University, Stanford, CA 94305, USA}};
\end{tikzpicture}
\end{equation*}
\begin{equation*}
\begin{tikzpicture}
\draw (0, -13.1) node[right, text width=0.45\paperwidth] {\texttt{scopelliti@lorentz.leidenuniv.nl, kschalm@lorentz.leidenuniv.nl, ajlucas@stanford.edu}};
\draw (\textwidth, -13.6) node[left] {\textsf{\today}};
\end{tikzpicture}
\end{equation*}
\begin{equation*}
\begin{tikzpicture}
\draw[very thick, color=\stylecolor] (0.0\textwidth, -5.75) -- (0.99\textwidth, -5.75);
\draw (0.12\textwidth, -6.25) node[left] {\color{\stylecolor}  \textsf{\textbf{Abstract:}}};
\draw (0.53\textwidth, -6) node[below, text width=0.8\textwidth, text justified] {\small We develop the theory of hydrodynamic charge and heat transport in strongly interacting quasi-relativistic 
systems on manifolds with inhomogeneous spatial curvature.   In solid-state physics, this is analogous to strain disorder in the underlying lattice.  In the hydrodynamic limit, we find that the thermal and electrical conductivities are dominated by viscous effects, and that the thermal conductivity is most sensitive to this disorder.
We compare the effects of inhomogeneity in the spatial metric to inhomogeneity in the chemical potential, and discuss the extent to which our hydrodynamic theory is relevant for experimentally realizable condensed matter systems, including suspended graphene at the Dirac point.};
\end{tikzpicture}
\end{equation*}

\tableofcontents

\titleformat{\section}
  {\gdef\sectionlabel{}
   \Large\bfseries\scshape}
  {\gdef\sectionlabel{\thesection }}{0pt}
  {\begin{tikzpicture}[remember picture,overlay]
	\draw (1, 0) node[right] {\color{\stylecolor} \textsf{#1}};
	\fill[color=\stylecolor] (0,-0.35) rectangle (0.7, 0.35);
	\draw (0.35, 0) node {\color{white} \textsf{\sectionlabel}};
       \end{tikzpicture}
  }
\titlespacing*{\section}{0pt}{15pt}{15pt}

\begin{equation*}
\begin{tikzpicture}
\draw[very thick, color=\stylecolor] (0.0\textwidth, -5.75) -- (0.99\textwidth, -5.75);
\end{tikzpicture}
\end{equation*}

\section{Introduction}
A theory of electrical and thermal transport % in strange metals
necessarily relies on a precise description of how translation symmetry is broken. In conventional weakly coupled quasi-particle theories, %it
most collisions of electrons are with impurities or phonons, and relax momentum.   In recent years, rapid progress towards a theory of transport which also accounts for momentum-conserving
electron-electron interactions has been made \cite{lucasrmp}.
One of the most useful tools that has arisen for understanding transport in this limit is hydrodynamics.   Hydrodynamics is the effective theory describing the relaxation of any interacting system to thermal equilibrium on long wavelengths.   Such a theory is suitable for any interacting metal where the disorder which breaks translation invariance
only varies on long wavelengths compared to the electron-electron scattering length \cite{andreev,dsz, lucas, polini, levitovhydro, lucas3, torre}.  Although this is a difficult regime to reach experimentally, this has now become possible \cite{crossno,bandurin,mackenzie, levitov1703} (see also \cite{molenkamp}).    A thorough understanding of the hydrodynamic regime of transport is certainly necessary as a ``solvable" limit of any more complete theory of transport \cite{seantocome}.  Hence, it is worthwhile to have a systematic understanding of hydrodynamic transport in a broad variety of systems.    

The purpose of this paper is to describe hydrodynamic transport on curved spaces.\footnote{Our formalism is relatively similar to the emergent ``hydrodynamic" formalism used to describe transport in strongly correlated systems described via the AdS/CMT correspondence \cite{lucas, donos1506, grozdanov, donos1507, grozdanov2}.   However, in most of these papers, the random spatial metric is an emergent phenomenon from the point of view of the bulk description of the field theory --- the exception is \cite{Banks:2016wdh}.   We emphasize that we are interested in scenarios where the inhomogeneous spatial metric is a physical effect.} In electronic materials, the presence of internal strain on a crystal lattice can be interpreted as an effective distortion to the induced spatial metric \cite{ciarlet}.   As the electronic charge-carrying degrees of freedom move in this inhomogeneous metric, our results will be relevant for strongly correlated systems  in inhomogeneously strained crystals.   Following \cite{lucas3}, we will focus on the relativistic hydrodynamic equations as a model for  transport in monolayer graphene in the hydrodynamic limit.   The techniques which we develop straightforwardly generalize to other hydrodynamic models.  

Recent experimental evidence \cite{crossno} indicates that electrons behave hydrodynamically in charge-neutral graphene. Collectively they behave as a Dirac fluid: a plasma of thermally excited electron and holes which is likely to be strongly interacting at `reasonable' temperatures $T\sim 100$ K \cite{vafek, schmalian, schmalian2}. Crucial to the observation of this Dirac fluid is the reduction and smoothing of ``charge puddle" disorder, which corresponds to inhomogeneities in the local chemical potential.   This was achieved by placing the graphene sheet in between layers of another material:  boron nitride \cite{xue}.   Another way to reduce charge puddle disorder in graphene is to ``suspend" graphene,  leaving it unattached to any substrate \cite{bolotin, mayorov}.   For mechanical reasons, dealing with such suspended graphene can be challenging.  The aspect we focus on here is that in principle a suspended sheet of graphene, as it consists of a single two dimensional `membrane' of carbon atoms, is susceptible to out-of-plane flexural distortions.   From the point of view of a two-dimensional effective theory for the Dirac fluid, flexural disorder can be interpreted as disorder in the spatial components of the spacetime metric.   Letting the local height of the membrane be $h(x,y)$, the metric is \cite{ciarlet} \begin{equation}
\mathrm{d}s^2 = \left(\mdelta_{ij} + \partial_i h \partial_j h \right) \mathrm{d}x_i \mathrm{d}x_j.  \label{eq:hmetric}
\end{equation}   
In reality, $h(x,y)$ need not be time-independent.  However, such flexural motion is expected to be quite slow relative to electronic time scales, and we may approximate it as static disorder.  Hence, a study of hydrodynamic electron transport in suspended graphene should naturally include  flexural distortions to the metric.

The outline of this paper, and our main conclusions, are as follows:  \begin{itemize}
\item In Section \ref{sec2}, we review the theory of linearized relativistic hydrodynamics on curved spaces, relevant for transport.

\item In Section \ref{sec3} we use this curved space hydrodynamics to solve for thermoelectric transport coefficients for a fluid in a slowly varying chemical potential and spatial metric.     When the inhomogeneity is small, we give analytic expressions for the thermal and charge conductivities as functions, expressed entirely in terms of the inhomogeneous chemical potential and metric, and thermodynamic and hydrodynamic coefficients.   When the inhomogeneity cannot be treated analytically, we compute the transport coefficients numerically.     Because transport is dissipative, the transport coefficients depend on hydrodynamic dissipation, via viscosity and a ``quantum critical" conductivity.   In the presence of inhomogeneous chemical potentials, both dissipative channels affect the conductivity significantly.   However, for inhomogeneous strain viscous dissipation is far more relevant -- in fact, perturbatively it is the \emph{only} source of dissipation.

\item We discuss the application of our formalism to suspended graphene in Section \ref{sec4}.  This discussion includes a justification of some of the statements in the introduction.   Our hydrodynamic transport theory allows us to describe electronic scattering off of certain long-wavelength phonons non-perturbatively in the strength of electronic interactions.    Although we will see that most phonons cannot be accounted for in this limit, our results may nonetheless be valuable for  a more detailed study of electron-phonon coupling.
\end{itemize}
Technical results are found in appendices.  We mostly work in units where $\hbar=k_{\mathrm{B}}=1$, and we also set the effective speed of light $v_{\mathrm{F}}=1$,\footnote{In materials such as graphene, the effective speed of light is set by the Fermi velocity $v_{\mathrm{F}}$.} as well as the electron charge $e=1$.    When we discuss the application of our formalism to suspended graphene, we will briefly restore these dimensionful quantities.

\section{Relativistic Hydrodynamics on Curved Space}
\label{sec2}

In this section we review and generalize to curved spacetime the hydrodynamic framework developed in \cite{lucas3}. This framework describes the collective motion of the relativistic electronic plasma in a disordered metal, where the disorder is introduced via a spatially dependent chemical potential $\mu_0(\mathbf{x})$.  When the chemical potential varies on a length scale larger than the electron mean-free path, a hydrodynamic description of transport is sensible: all other microscopic degrees of freedom have already reached local thermodynamic equilibrium.   The only relevant degrees of freedom for transport are locally conserved quantities:  energy, charge and momentum.   All the spatial dependence of the parameters (such as local energy density $\epsilon$ or shear viscosity $\eta$) is encoded by the functional dependence of these quantities on the local $\mu_0(\mathbf{x})$:  e.g. $\eta(\mathbf{x}) = \eta(\mu_0(\mathbf{x}))$.
% The foundation of this framework is the recognition that if disorder is mediated through a conserved current, the thermo-electric response is universal \cite{andreev,dsz}. 
Charge/chemical potential disorder is natural for many metals,  including graphene \cite{lucas3}.  For slowly varying disorder, this is also convenient because it is very naturally included within a hydrodynamic framework.

Another type of universal disorder that is natural to consider within a hydrodynamic framework is local inhomogeneity in the spacetime metric:  as we described previously, this is a model for strain in the crystal lattice.    This strain can also be natural in a broad variety of solids:  occuring from either in-plane strain, or (in the case of suspended graphene) out-of-plane bending of the crystal lattice.  In the limit where this strain is long wavelength, we can account for it by simply solving the hydrodynamic equations of motion, written in a coordinate-independent fashion, on a curved spacetime.   

 Let us note that strain can also open up a gap  $\Delta$ in certain crystals, including graphene \cite{zhni}.    This will alter the effective microscopic dispersion relation and hence the equations of state.   In the present work we have neglected this contribution, and our theory is not valid if the strain is so large that $\Delta \sim T$.    For smaller strain, our theory remains valid, but there will be additional $\mathbf{x}$-dependence of the thermodynamic and hydrodynamic coefficients due to the local value of the gap.   For simplicity we will not explicitly acocunt for this effect.   Up to the opening of a gap,  the effects of strain are universal in the hydrodynamic limit.

As we previously noted, the only quantities which are globally conserved (up to external sources) are charge, energy and momentum \cite{kovtun}.   The natural degrees of freedom are the local number density $n(x)$, the energy density $\epsilon(x)$ and the momentum density $\Pi_i(x)$.   A more convenient approach is to use their thermodynamic conjugates: the chemical potential  $\mu(x)$, temperature  $T(x)$, and velocity  $v^{i}(x)$ respectively.   These are the standard choice of hydrodynamic variables.  In relativistic systems this velocity is commonly written covariantly as a four-velocity $u^{\mu}(x)=(1, v^i)/\sqrt{1-v^2}$, constrained to equal $u^{\mu}u^{\nu}\eta_{\mu\nu}=-1$ with $\eta_{\mu\nu}$ the Minkowski metric:  $\eta_{\mu\nu} = \mathrm{diag}(-1,1,\ldots,1)$.     

The equations of motion are local conservation laws:
\begin{subequations}\label{eq:hydroEqu}\begin{align}
&\nabla_{\mu}T^{\mu\nu}=F_{\mathrm{ext}}^{\nu\mu}J_{\mu},\label{eq:1}\\
&\nabla_{\mu}J^{\mu}=0 \label{eq:2}.
\end{align}
\end{subequations} 
In the absence of an external electric field or temperature gradient, there remains an external electromagnetic field due to an inhomogeneous chemical potential:  $F^{\mu\nu}_{\mathrm{ext}} = \nabla^\mu A^\nu_{\mathrm{ext}} - \nabla^\nu A^\mu_{\mathrm{ext}}$, with \begin{equation} 
A_{\mathrm{ext}} =  \mu_0(\mathbf{x}) \mathrm{d}t.
\end{equation}
The only non vanishing components of the Maxwell tensor are $F_{ti}^{\mathrm{ext}} = -F_{it}^{\mathrm{ext}} = \partial_i \mu_0(\mathbf{x})$.    $T^{\mu\nu}(\mathbf{x},t)$ and $J^\mu (\mathbf{x},t)$ are the expectation values of the local relativistic stress-energy tensor and charge current, respectively. These conservation equations, understood in terms of the covariant derivative $\nabla_\mu$ with respect to the metric \begin{equation}
\mathrm{d}s^2 = g_{\mu\nu}\mathrm{d}x^{\mu}\mathrm{d}x^{\nu} = -\mathrm{d}t^2 + g_{ij}(\mathbf{x}) \mathrm{d}x_i \mathrm{d}x_j,
\end{equation} are valid in any (curved) spacetime, including those with an inhomogenous spatial curvature of interest to us.

In order for the equations (\ref{eq:hydroEqu}) to be well-posed, we must express the expectation value of $T^{\mu\nu}$ and $J^{\mu}$ in terms of the hydrodynamic variables $\mu$, $T$ and $u^\mu$. %In Section \ref{sec:App1.1} we show that there is a static solution with $u^\mu = (1,0,0)$, $T = T_0 =$ constant and $\mu(\mathbf{x})=\mu_0(\mathbf{x})$.\\   
We will expand $T^{\mu\nu}$ and $J^\mu$ in a gradient expansion in derivatives:  more physically, the small parameter of the perturbative expansion is $\ell_{\mathrm{ee}} k$, with $k$ the wave number of our perturbation, and $\ell_{\mathrm{ee}}$ the electron-electron scattering length.  In this paper, we will  only include terms with zero or one derivatives of $\mathbf{x}$ and $t$.   This expansion is well-known for a relativistic fluid \cite{kovtun,hkms}:
\begin{subequations}
\label{eq:hydroExpans}
\begin{align}
T^{\mu\nu}&=(\epsilon+P)u^{\mu}u^{\nu}+P g^{\mu\nu}-2\eta\mathcal{P}^{\mu\rho} \mathcal{P}^{\nu\sigma}\nabla_{(\rho}u_{\sigma)}-\mathcal{P}^{\mu\nu}\left(\zeta-\frac{2\eta}{d}\right)\nabla_{\rho}u^{\rho},	\label{eq:stressEn}\\
J^{\mu}&= nu^{\mu}-{\sigma_{\textsc{q}}}\mathcal{P}^{\mu\nu}\left(\partial_{\mu}\mu-\frac{\mu}{T}\partial_{\nu}T-F_{\nu\rho,\mathrm{ext}}u^{\rho}\right),		\label{eq:electCurr}
\end{align}\end{subequations}
with $\eta$ and $\zeta$ the shear and bulk viscosity respectively, and $\sigma_{\textsc{q}}$ a microscopic dissipative coefficient.  As emphasized in \cite{hkms},   $\sigma_{\textsc{q}}$ should be interpreted as the \emph{finite} electrical conductivity of the charge neutral plasma (up to hydrodynamic long-time tails \cite{lucasrmp}), and for historical reasons it is sometimes called the ``quantum critical" conductivity.    Finally, $\mathcal{P}^{\mu\nu}$ is the projector orthogonal to the rest frame of the fluid, set by the velocity $u^\mu$:   $\mathcal{P}^{\mu\nu}= g^{\mu\nu}+u^\mu u^\nu$.

\section{Hydrodynamic Transport}
\label{sec3}

We now wish to compute the thermoelectric conductivity matrix of a fluid in such an inhomogeneous background.  These coefficients are defined as follows: \begin{equation}
\left(\begin{array}{c} J_i^{\mathrm{avg}} \\ Q_i^{\mathrm{avg}}  \end{array}\right) \equiv \left(\begin{array}{cc}  \sigma_{ij} &\  T\alpha_{ij} \\ T\bar\alpha_{ij} &\ T\bar\kappa_{ij} \end{array}\right)  \left(\begin{array}{c} E_j \\ \zeta_j  \end{array}\right),
\end{equation}
where $J_i^{\mathrm{avg}}$ is the spatial average of the charge current defined above,  $Q_i^{\mathrm{avg}}$ is the spatial average of the heat current, defined as 
\begin{equation}
Q^i \equiv T^{ti} - \mu(\mathbf{x})J^i,
\end{equation}
$E_j$ is an infinitesimal externally applied uniform electric field,  and $\zeta_j $ is an infinitesimal ``thermal drive" analogous to a homogeneous temperature gradient $-\partial_j \log T$.   This more formal notation will prove useful for our purposes.

Our goal is to compute $\sigma_{ij}$,  $\alpha_{ij}$, $\bar\alpha_{ij}$ and $\bar\kappa_{ij}$, using the hydrodynamic  equations of motion.  We will explicitly show how this is done.   First, let us note a few formal results.   Onsager reciprocity states that (with time-reversal symmetry)  $\bar\alpha_{ij} = \alpha_{ji}$, and that $\sigma$ and $\bar\kappa$ are symmetric.   In the hydrodynamic framework on a curved space, we prove this in Appendix \ref{sec:Onsager}.   Secondly, it is experimentally more common to measure a thermal conductivity defined by \begin{equation}
\kappa_{ij} \equiv \bar\kappa_{ij} - T\bar\alpha_{ik}\sigma^{-1}_{kl}\alpha_{lj}.
\end{equation}
This can be interpreted as the ratio of the average heat current to a constant temperature gradient, subject to the constraint that no net charge current flows.   We will show results for $\bar\kappa$ and  for $\kappa$.

\subsection{General Solution}

%As discussed in the introduction, weak universal disorder is encoded by allowing the parameters to become position dependent in response to a slowly varying external chemical potential $\mu_{0} \rightarrow \mu_{0}(\mathbf{x})$ and a slowly varying spatial metric $g_{00}\rightarrow -1, g_{0i} \rightarrow 0, g_{ij} \rightarrow g_{ij}(\mathbf{x})$. 
We now present the formal computation of the thermoelectric conductivity matrix.    First, we note that in an inhomogeneous metric $g_{ij}(\mathbf{x})$ and chemical potential $\mu_0(\mathbf{x})$, there is an exact solution to the nonlinear equations of motion, encoding that the fluid is at rest in local thermal equilibrium:  \begin{subequations}\begin{align}
\mu_{\mathrm{eq}}(\mathbf{x}) &= \mu_0(\mathbf{x}), \\
T_{\mathrm{eq}}(\mathbf{x}) &= T_0, \\
u^\mu_{\mathrm{eq}}(\mathbf{x}) &= (1,\mathbf{0}),
\end{align}\end{subequations}
where $T_0$ is a constant.     Then, because we are applying an infinitesimal electric field and thermal drive, we only look for the perturbations around equilibrium within linear response:  \begin{subequations}\begin{align}
\mu(\mathbf{x}) &\approx \mu_{\mathrm{eq}}(\mathbf{x}) + \mdelta \mu(\mathbf{x}), \\
T(\mathbf{x})  &\approx   T_{\mathrm{eq}}(\mathbf{x}) + \mdelta T(\mathbf{x}), \\
u^\mu(\mathbf{x}) &\approx (1, \mdelta v^i(\mathbf{x})).
\end{align}\end{subequations}

Because the disorder explicitly picks out a preferred fluid rest frame, it is often helpful to decompose (\ref{eq:hydroEqu}) into timelike and spacelike components.    The hydrodynamic expansion of the electric current within linear response gives:
\begin{subequations}
\begin{align}
J^t &= n,\\
J^j &= n\mdelta v^j - {\sigma_{\textsc{q}}} g^{ij}\left(\partial_{i}\mdelta \mu - \frac{\mu_0}{T_0} \partial_{i}\mdelta T \right),
\end{align}
\end{subequations}
\noindent while the stress-energy tensor reads % \eqref{eq:stressEn}
\begin{subequations}
\begin{align}
T^{tt}&=\epsilon,\\
T^{ti}&= (\epsilon+P)\mdelta v^{i}% + g^{ij} \zeta_{j}P,
  \\
%T^{0}_{\,\,\,i}&=\frac{\zeta_{i}}{i\omega}e^{-i\omega t}(\epsilon + P) +(\epsilon + P) g_{ij}v^j ,\\
T^{ij}&=(P_0 + \mdelta P) g^{ij}-\eta\left(\bar{\nabla}^{j}\mdelta v^{i} + \bar{\nabla}^{i}\mdelta v^{j} \right)-\left(\zeta-\frac{2}{d}\eta\right)g^{ij}\bar{\nabla}_{k}\mdelta v^{k},
%T^{i}_{\,\,\,j}&= P\delta^{i}_{\,\,\,j}-\eta\left( \bar{\nabla}_{j}v^{i} + \bar{\nabla}^{i}v_{j}\right)-\left(\zeta-\eta\right)\delta^{i}_{\,\,\,j} \bar{\nabla}_{k}v^{k}.
\end{align}
\end{subequations}
where $\bar{\nabla}_i v^j \equiv \partial_iv^j + \Gamma^{j}_{~kl}v^j$ is the covariant derivative with respect to the spatial metric $g_{ij}$ and  $\Gamma^{j}_{~kl}=\frac{1}{2}g^{jm}(\partial_k g_{ml}+\partial_lg_{mk}-\partial_m g_{kl})$ is the Christoffel symbol.   For simplicity, we henceforth  specialize to two spatial dimensions: $d=2$.

The external electric field $E_i$ and thermal drive $\zeta_i$ are added by modifying the background vector potential $A$ and spacetime metric $g$ \cite{lucasrmp}:  \begin{subequations}
\begin{align}
A &= \mu_0(\mathbf{x}) \mathrm{d}t + (E_i - \mu(\mathbf{x}) \zeta_i) \frac{\mathrm{e}^{-\mathrm{i}\omega t}}{\mathrm{i}\omega} \mathrm{d}x_i, \\
&\mathrm{d}s^{2}=-\mathrm{d}t^2+g_{ij}(\mathbf{x})\mathrm{d}x^{i}\mathrm{d}x^{j}+2\frac{\mathrm{e}^{-\mathrm{i}\omega t}}{\mathrm{i}\omega}\zeta_{i}\mathrm{d}x^{i}\mathrm{d}t.
 \label{eq:metrDeform}
\end{align}
\end{subequations}

We are interested in the thermoelectric conductivities within linear response, and so we need only calculate the perturbations $\mdelta \mu$, $\mdelta T$ and $\mdelta v_i$ to linear order in $E_i$ and $\zeta_i$.   After some algebra, the linearized hydrodynamic equations can be found:
\begin{subequations}\label{eq:EOM}
\begin{align}
-\bar{\nabla}^{i}\left(\sigma_{\textsc{q}} \partial_{i}\mdelta\mu\right) +  \bar{\nabla}^{i}\left(\sigma_{\textsc{q}} \frac{\mu_{0}}{T_{0}}\partial_{i}\mdelta T\right) + \bar{\nabla}_i \left(n\mdelta v^i\right)&= -\bar{\nabla}^{i}\left(\sigma_{\textsc{q}}\left(E_{i} - \mu_{0}\zeta_{i}\right) \right)\\
\bar{\nabla}^{i}\left(\sigma_{\textsc{q}} \mu_{0}\partial_{i}\mdelta\mu\right)  -\bar{\nabla}^{i} \left(\sigma_{\textsc{q}} \frac{\mu^{2}_{0}}{T_{0}}\partial_{i}\mdelta T\right) + \bar{\nabla}_i\left( s T_{0} \mdelta v^i\right)&=\bar{\nabla}^{i}\left(\sigma_{\textsc{q}}\mu_{0}\left(E_{i} - \mu_{0}\zeta_{i}\right)\right)\\
n \partial_{j}\mdelta\mu + s\partial_{j}\mdelta T  - \bar{\nabla}_i \left[\eta (\bar{\nabla}^{i}\mdelta v_j+ \bar{\nabla}_{j}\mdelta v^i)\right] - \partial_{j}\left[(\zeta - \eta)\bar{\nabla}_{i}\mdelta v^i\right]&=nE_{j}+sT_{0}\zeta_{j}
\end{align}
\end{subequations}
where $\bar{\nabla}$ is the covariant derivative with respect to the spatial component of the metric $g_{ij}$.     These are elliptic differential equations which can be straightforwardly solved numerically, as we describe in Appendix \ref{app:num}.

\subsection{Perturbative Analytic Solution}
In the limit where $g_{ij}$ % can be expressed in the form (\ref{eq:hmetric}), with $h(\mathbf{x})$ perturbatively small
is a perturbatively small deviation from flat space  $g_{ij}=\delta_{ij}+\hat{g}_{ij}$, and the spatial variation of the chemical potential around the average $\mu_0(\mathbf{x}) =\bar{\mu}_0+\hat{\mu}(\mathbf{x})$ is also perturbatively small,  we can analytically compute the conductivity matrix to leading order.    The calculation is rather tedious and is presented in Appendix \ref{sec:weakDis}.\footnote{Note that $\hat{g}_{ij}$ is quadratic in the  height function of out-of-plane distortions.   For a chemical potential and induced metric with an explicit small parameter $u$,  $\mu_0(\mathbf{x}) = \bar\mu_0 + u \hat\mu(\mathbf{x})$ and $ g_{ij} = \mdelta_{ij} + u \partial_i \hat h \partial_j \hat h$, with $\hat \mu$ and $\hat h$ O(1) functions.  Hence fluctuations in the height function $\sqrt{u}\hat h$ must be parametrically larger amplitude than the fluctuations in the chemical potential, $u\hat \mu$, to have the same effect on transport.      When $u\rightarrow 0$, the transport coefficients $\sigma_{ij}$, $\alpha_{ij}$ and $\bar\kappa_{ij}$ will be $\mathrm{O}(u^{-2})$.}   The transport coefficients can be expressed in terms of the relaxation rate $\tau^{-1}_{ij}$ for momentum.   Assuming the density $n$ is finite, one expects on general grounds \cite{lucasrmp}:
\begin{subequations}
\label{eq:TranspMain}
	\begin{align}
	&\sigma^{ij}\approx \frac{ n^2 \tau^{ij}}{\epsilon + P},\\
	&\alpha^{ij}\approx \frac{n s \tau^{ij}}{\epsilon + P},\\
	&\bar{\kappa}^{ij}\approx \frac{T s^2 \tau^{ij}}{\epsilon + P},
	\end{align}
\end{subequations}
%This small disorder inverse scattering rate $\tau_{ij}^{-1}$ receives contributions from pure charge disorder, pure metric disorder and the interaction between metric and charge disorder. This last contribution, however, will be not relevant in the thermodynamic limit, since the disorder distributions self-average $\left\langle\delta\mu(\mathbf{k})\delta\mu(\mathbf{k}')\right\rangle\sim \delta((\mathbf{k-k}'))$,~$\left\langle h(\mathbf{k})h(\mathbf{k}')\right\rangle\sim \delta((\mathbf{k-k}'))$ and then the metric and charge disorder are not correlated $\left\langle\delta\mu(\mathbf{k})h(\mathbf{k}')\right\rangle\sim 0$.
%\comment{[I moved the footnote to main text]}. The separate disorder contributions add following Matthiessen's rule as usual for single degree-of-freedom transport and they are
where \begin{equation}
\tau_{ij}^{-1} = (\tau_{ij}^{-1})^{(\mu\mu)} + (\tau_{ij}^{-1})^{(\mu h)}+ (\tau_{ij}^{-1})^{(h h)}
\end{equation}
with
\begin{subequations}
\label{eq:scatteringRate}
\begin{align}
(\tau_{ij}^{-1})^{(\mu\mu)} =&  \sum_{\mathbf{k}} \frac{k_i k_j}{k^2} \frac{\left|T_0 n_0 \hat s(\mathbf{k}) - T_0 s_0 \hat n(\mathbf{k})\right|^2 + k^2 \sigma_{\textsc{q}}(\eta_0+\zeta_0) \left|T_0 \hat s(\mathbf{k}) + \mu_0 \hat n(\mathbf{k})\right|^2}{\sigma_{\textsc{q}}(\epsilon_0 + P_0)^3} \\
(\tau_{ij}^{-1})^{(\mu h)} =& 2\eta_0  \sum_{\mathbf{k}}  k_i k_j \, \frac{\bar\mu_0 n(\mathbf{k})+T_0 s(\mathbf{k})}{(\epsilon_0 + P_0)^2} \hat{g}_{kl}(-\mathbf{k})\, \mathbb{P}_{kl} 
 \\
(\tau_{ij}^{-1})^{(h h)} =& \frac{\eta_0}{\epsilon_0 + P_0} \sum_{\mathbf{k}}  k_{i} k_j  \,  \hat{g}_{rs}(\mathbf{k})\hat{g}_{kl}(-\mathbf{k}) \mathbb{P}_{r(s} \mathbb{P}_{k)l}.
\end{align}\end{subequations}
We have defined the projector
\begin{equation}
\mathbb{P}_{ij}=\mdelta_{ij}-\frac{k_i k_j}{k^2}.  \label{eq:proj}
\end{equation} 
The pure charge disorder scattering rate $(\tau_{ij}^{-1})^{(\mu\mu)}$ was found before in \cite{lucas3}.  If the disorder in the chemical potential is uncorrelated with the strain disorder, then after disorder averaging we expect $(\tau^{-1})^{(\mu h)} \approx 0$. 

As we have stated, $\tau^{-1}_{ij}$ is the rate at which the fluid can relax its momentum on the long wavelength disorder.   As this is a dissipative process, it is necessarily the case that $\tau^{-1}_{ij}$ must depend on $\sigma_{\textsc{q}}$, $\eta$ and/or $\zeta$.    As we show in Appendix \ref{sec:weakDis}, in this perturbative regime the charge and heat currents are -- at leading order -- uniform.  When there are inhomogeneities in the chemical potential, momentum relaxation can be non-negligible, even in the limit where inhomogeneity is very long-wavelength ($k\rightarrow 0$).  This is due to the fact that a uniform fluid velocity, uniform charge and uniform heat current are not simultaneously consistent with both charge and heat conservation -- the heat and charge currents must contain a purely dissipative component, which carries no momentum.  The conductivity associated with this incoherent current is proportional to $\sigma_{\textsc{q}}$;  this explains the $1/\sigma_{\textsc{q}}$ scaling of the first term in $(\tau^{-1})^{(\mu\mu)}$.   A non-relativistic avatar of this effect was emphasized in \cite{andreev}.    However, in the presence of strain or metric disorder, there is no such impediment to a uniform flow.   In this case, dissipation arises due to viscous effects.   As viscosity vanishes, dissipation becomes weaker and hence the thermoelectric conductivity matrix is proportional to $1/\eta$.   

%Note that this pure charge disorder scattering vanishes if the microscopic conductivity $\sigma_{\textsc{q}}$ vanishes. Pure charge disorder does not generically lead to momentum dissipation as in most physical systems $\sigma_{\textsc{q}}$ is negligibly small. \comment{[CHECK]}

\begin{figure}[h]
\pgfplotsset{every axis legend/.append style={at={(.03,0.99)},anchor=north west}}
\pgfplotsset{title style={at={(0.3,0.8)},anchor=north west}}
	\centering
	\includegraphics[scale=0.92]{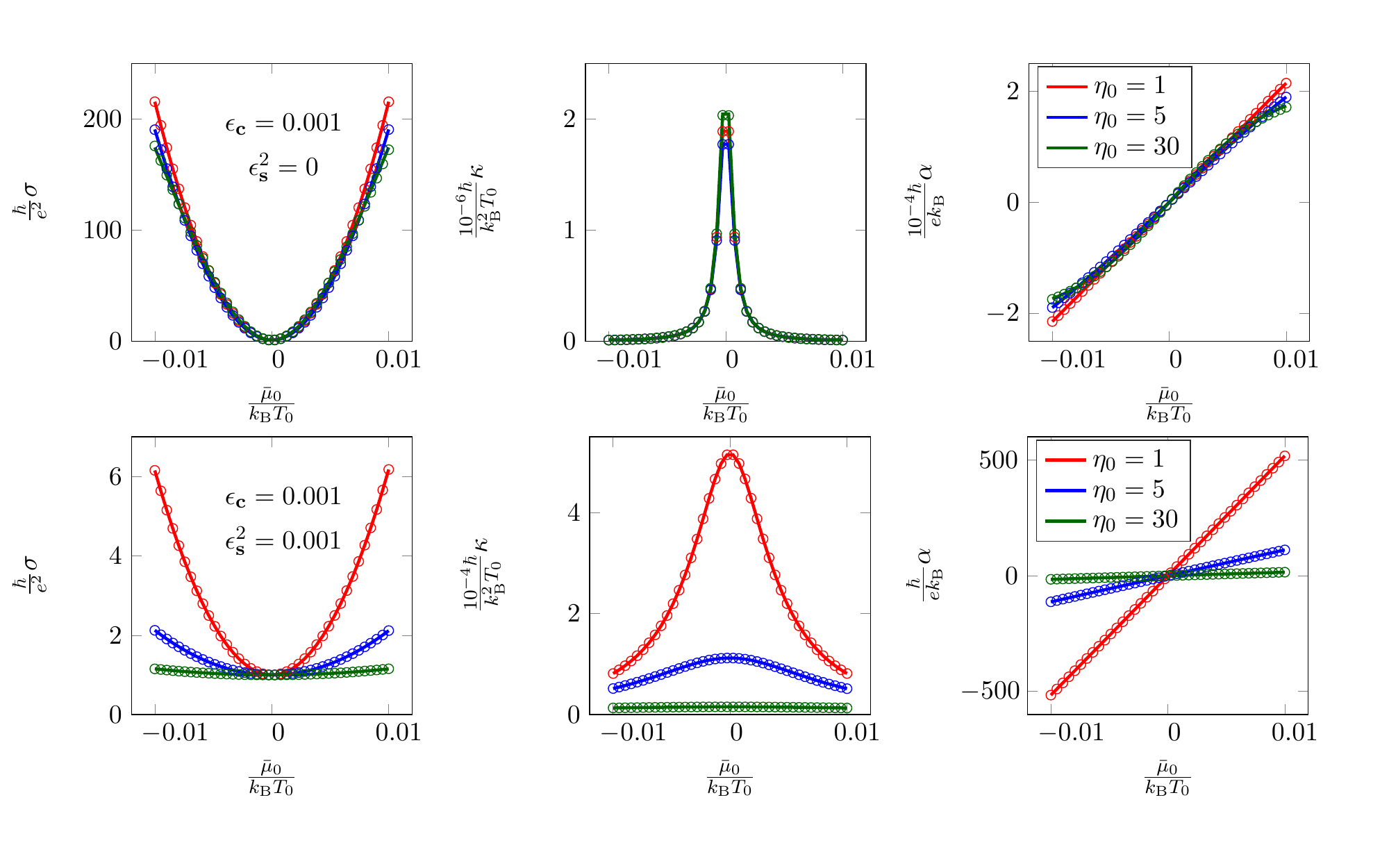}
	\caption{Numerical simulation of the transport coefficients in dimensionless units for weak disorder with $C_0 = C_2 =\sigma_0 =1$.    For convenience in all of our figures, we have restored dimensional prefactors of $\hbar$, $e$, $T_0$ and $k_{\mathrm{B}}$.   Numerical results (circular markers) agree very well with  the theoretical results \eqref{eq:TranspMain} and (\ref{eq:scatteringRate}) (solid lines). In the first row, only charge disorder is present and the dependence on the shear viscosity is very weak. Switching on strain disorder considerably increases  the sensitivity to shear viscosity $\eta$. The results have been averaged over 20 disorder configurations.}
\label{fig:numerics1}
\end{figure}

\subsection{Numerical Solution}
\label{npt}
In order to make more precise predictions, the theory introduced in the previous section ought to be supplemented with specific equations of state.   As noted in \cite{lucas3}, the equations of state for a quasirelativistic fluid with gapless excitations, such as the Dirac fluid in graphene, are rather constrained.   If we focus on the physics near the charge neutrality point for simplicity,  we find:
\begin{subequations} \label{eq:EOS}
\begin{align}
&n(\mu_0)=C_2 \mu T+ \mathrm{O}\left(\frac{\mu^3}{T}\right),\\
&s(\mu_0)= C_0 T^2 + \frac{C_2}{2}\mu^2  + \mathrm{O}\left(\frac{\mu^4}{T^2}\right),\\
&\eta(\mu_0)=T^2 \eta_0 + \mathrm{O}\left(\mu^2\right),\\
&\zeta(\mu_0)=0,\\
&\sigma(\mu_0)=\sigma_0 + \mathrm{O}\left(\frac{\mu^2}{T^2}\right),
\end{align}
\end{subequations}
 where the constants $\sigma_0$, $\eta_0$ and $C_{0,2}$ are dimensionless.    For simplicity we have assumed that the bulk viscosity $\zeta=0$; we did not find that a finite $\zeta$ led to qualitatively different physics than a finite $\eta$.

   Using the spectral methods of \cite{lucas3}, described in Appendix \ref{app:num}, we have numerically solved (\ref{eq:EOM}) with the equations of state (\ref{eq:EOS}), in inhomogeneous chemical potentials and metrics.   We have always taken periodic boundary conditions, and assumed that the metric disorder and chemical potential disorder are uncorrelated, for simplicity.

Denoting spatial averages with $\mathbb{E}[\cdots]$,  let us define \begin{subequations}\begin{align}
\epsilon_{\textsc{c}}^2 &= T^{-2} \mathbb{E}\left[(\mu(\mathbf{x})-\bar\mu_0)^2\right], \\
\epsilon_{\textsc{s}}^2 &= T^2 \mathbb{E}\left[h(\mathbf{x})^2\right].
\end{align}\end{subequations} 
These two parameters quantify the relative amount of charge vs. strain disorder.   The overall prefactors of temperature $T$ are chosen so that $\epsilon_{\textsc{c,s}}$ are dimensionless numbers.    
In Figure \ref{fig:numerics1}, we  demonstrate % that the numerical methods agree with the analytic results quite precisely when $\epsilon_{\textsc{c,s}} \ll 1$.   This figure also makes
quite clearly the dramatic effects of viscosity on transport in the presence of strain disorder, as explained in the previous section. 

\begin{figure}[h]
	\centering
	\includegraphics[scale=1]{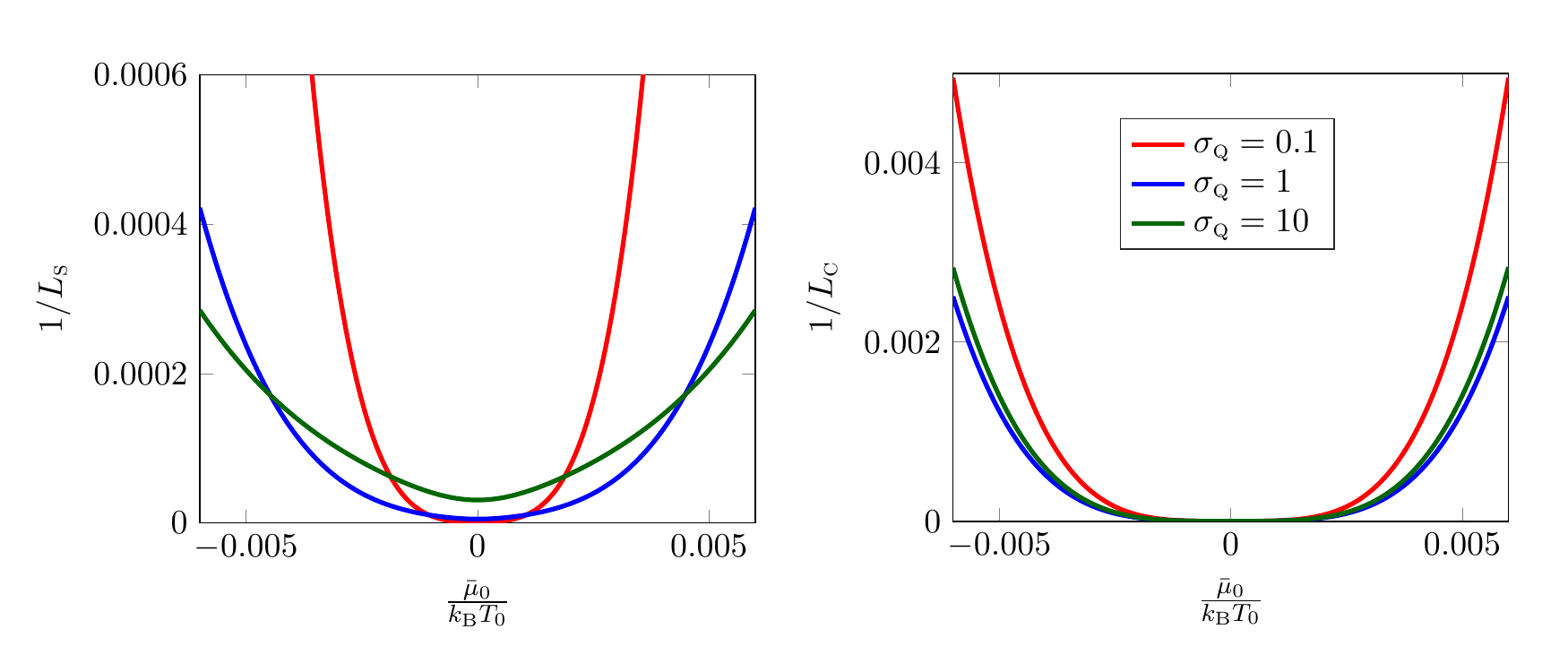}
	\caption{Numerical simulation of the Lorenz ratio $L$, with $C_0 = C_2 =1$ and $\eta_0 =5$, with variable $\sigma_{\textsc{q}}$.  Left plot: only strain disorder   ($\epsilon_{\textsc{c}} = 0$ and $\epsilon_{\textsc{s}}^2=0.001$);
 right plot:  only charge disorder ($\epsilon_{\textsc{c}} = 0.001$).   As expected, the Lorenz ratio is much more sensitive to $\sigma_{\textsc{q}}$ with strain disorder, relative to charge disorder.}
\label{relativenumerics}
\end{figure}

The other dissipative channel is the one controlled by the microscopic conductivity $\sigma_{\textsc{q}}$. The presence of $\sigma_{\textsc{q}}$ is essential: for vanishing $\sigma_{\textsc{q}}=0$ there can be no heat current in the absence of an electric current, and so $\kappa=0$.   So a clear way to observe the effects of $\sigma_{\textsc{q}}$ is in the Lorenz ratio \begin{equation}
L = \frac{\kappa}{T\sigma},
\end{equation}
where for simplicity we have assumed isotropic transport coefficients (this is the case for isotropic disorder).   For perturbatively small disorder, we estimate the Lorenz ratio \cite{lucas3} \begin{equation}
\kappa \approx \left\lbrace \begin{array}{cc} \displaystyle \dfrac{(\epsilon+P)\tau}{T^2\sigma_{\textsc{q}}} &\ n_0 \approx 0 \\ \dfrac{(\epsilon+P)^3 \sigma_{\textsc{q}}}{T^2 n^4 \tau }  &\  \text{otherwise}  \end{array}\right.
\end{equation} as can be seen from the analytic results.   For chemical potential disorder, we expect that (at small $\sigma_{\textsc{q}}$)  $\tau \sim \sigma_{\textsc{q}}$ and so $\kappa$ does not depend strongly on $\sigma_{\textsc{q}}$.   When all disorder is in the strain,  $\tau$ does not depend on $\sigma_{\textsc{q}}$ and so $L$ has much stronger dependence on $\sigma_{\textsc{q}}$.  This is shown in Figure \ref{relativenumerics}.

\begin{figure}
\pgfplotsset{every axis legend/.append style={at={(.03,0.99)},anchor=north west}}
\pgfplotsset{title style={at={(0.3,0.8)},anchor=north west}}
	\centering
	\includegraphics[scale=0.87]{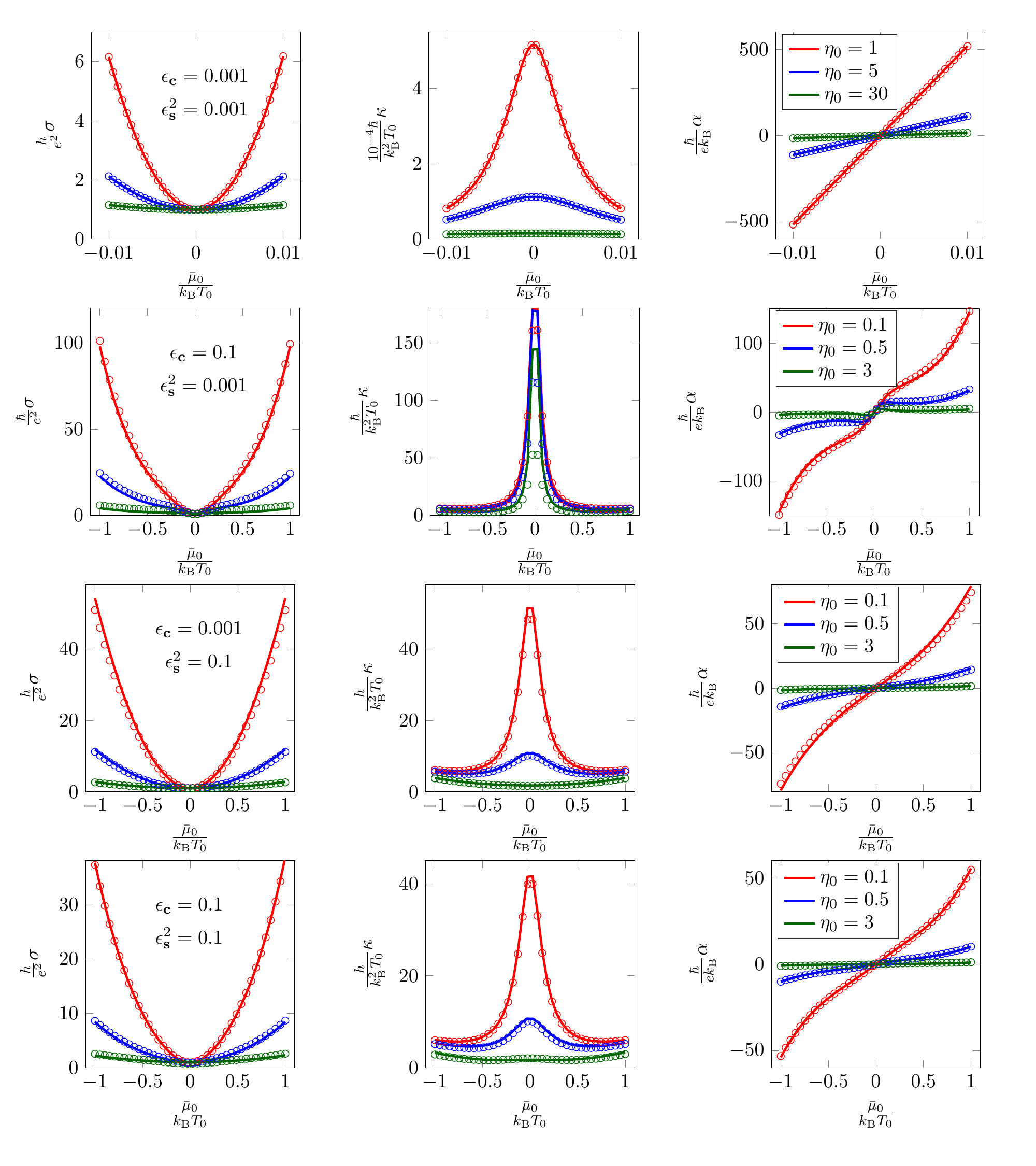}
	\caption{Numerical simulation of the transport coefficients in dimensionless units with $C_0 = C_2 =\sigma_0 =1$.   The circular markers represent the numerical results, while the solid lines represent the theoretical results \eqref{eq:TranspMain} and (\ref{eq:scatteringRate}). The agreement between numerics and analytics decreases upon increasing the  strength of disorder, although the agreement remains better for larger strain disorder vs. chemical potential disorder.   The results have been averaged over 20 disorder configurations.}
\label{fig:numerics2}
\end{figure}

The numerical results in Figure \ref{fig:numerics1} and Figure \ref{relativenumerics} are still fully in the perturbative analytic regime.  For larger disorder the analytic results are no longer quantitatively correct, even though the differences remain small and the qualitative features stay the same.  This is shown in Figure  \ref{fig:numerics2}. 

\begin{figure}
	\centering
	\includegraphics[scale=1]{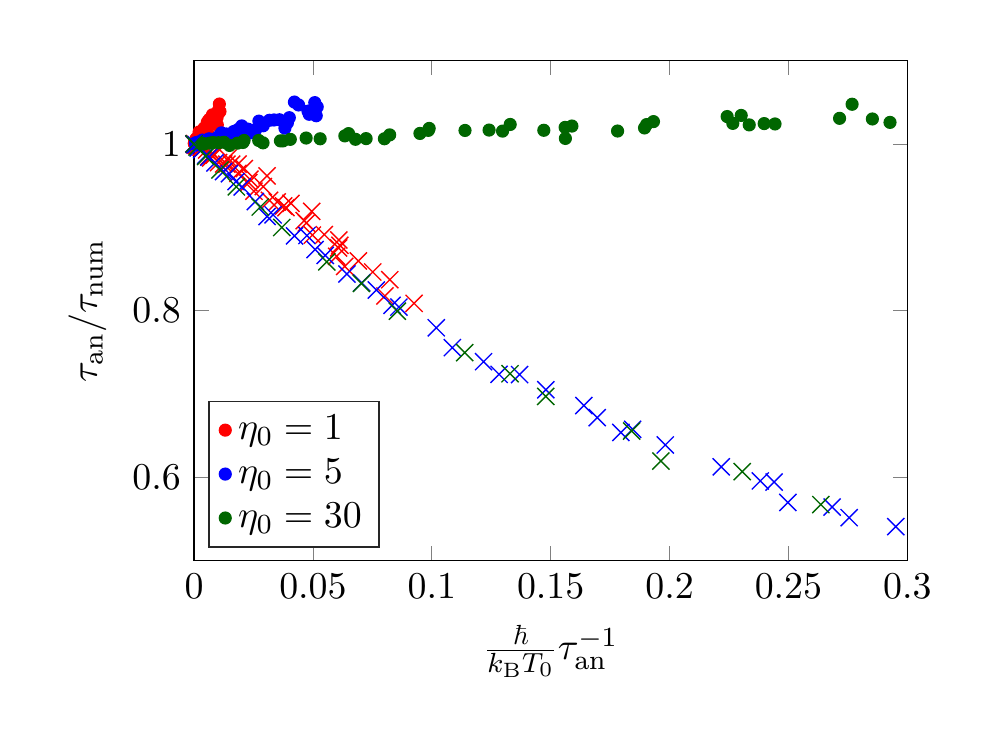}
	\caption{We analyze the validity of the perturbative solution by comparing the ratio of the numerically estimated scattering rate $\tau_{\text{num}} = (\epsilon+P)\sigma/n^2$ to our analytic prediction $\tau_{\text{an}}$ (\ref{eq:scatteringRate}).  The dots represent the ratio of scattering rates in presence of pure strain disorder with $\epsilon_{\textsc{s}}^2$ in the interval $[0.01,0.1]$. The crosses represent pure charge disorder with $\epsilon_{\textsc{c}}$ in the interval $[0.001, 0.1]$.   We have set $\bar\mu_0/T=2$.   In the perturbative regime where $\tau^{-1}\rightarrow 0$ the analytic and numerical results match, as they must.}
\label{fig:non-pert}
\end{figure}

A clear indication that one is outside the perturbative regime is that the results can no longer be described in terms of a sum of inverse scattering times.  This is depicted in Figure \ref{fig:non-pert}.  Beyond the perturbative regime, we find that the analytic expression \emph{overestimates} the conductivity in the presence of strain disorder, and \emph{underestimates} the conductivity in the presence of charge disorder.

% \new{Still, many qualitative dependencies persist. From the perturbative analytic result, it is seen that the shear viscosity controls the strength with which the strain disorder affects transport. Figure  \ref{relativenumerics} displays the relative strength of pure strain disorder to pure charge disorder transport and shows that this remains so for intermediate values of disorder.
% There are two qualitative regimes. For large chemical potential both the charge and thermal conductivity decrease with increasing shear viscosity. For small chemical potential $\mu \ll T$, the thermal conductivity retains this, but the charge conductivity now increases with increasing viscosity.  [AL:  IM NOT SURE WHAT THIS IS TRYING TO SAY?  LAST 2 SENTENCES CONFUSING.   ALSO 3 of the 4 PLOTS IN FIGURE 4 SEEM IDENTICAL, COULD JUST INCLUDE 1]   }

%\noindent Since all the transport is lead by the inverse scattering rate, we can reduce the investigation to the behavior of the electric conductivity only. From \eqref{eq:scattRate}, putting to zero the charge disorder (fluctuations), we see how the scattering rate for large spatial deformation decouples from the charge density and turns to be constant over $\mu_0$, as can be checked in figure \ref{fig:B}. Conversely, for small spatial deformation, the scattering rate still depends on the charge fluctuations and it cannot be approximated with a constant over $\mu_0$ CHECK THIS COMMENT

\section{Application to Suspended Graphene}
\label{sec4}

We now turn to the application of our formalism to hydrodynamics in the Dirac fluid in monolayer graphene \cite{vafek, schmalian}.    Graphene is a honeycomb lattice of carbon atoms in two spatial dimensions, with a low energy dispersion relation \begin{equation}
\epsilon_a(\mathbf{k}) = \hbar v_{\mathrm{F}}|\mathbf{k}|.
\end{equation}
The $a$ label denotes spin and valley indices, and will mostly be ignored for the purposes of this paper -- neither the interactions nor the disorder couples to spin here.    These electrons interact with one another via long-range Coulomb interactions.    Thus, strictly speaking,  the hydrodynamics of graphene cannot be relativistic hydrodynamics.

However, as we have seen, transport is a linear response calculation.   The key input from relativistic hydrodynamics was that the energy current and momentum density were identical -- this reduced the number of hydrodynamic variables present.   This follows  trivially from the (weak-coupling) action for the Dirac fluid, and so we expect that the non-relativistic nature of the interactions will not play an important role in a transport calculation.   Furthermore, as one can show following \cite{muller1, lucas3},  the effect of Coulomb interactions can be absorbed into a (nonlocal) redefinition of $\mu_{\mathrm{eq}}$ and $\mdelta \mu$, and so the final equations governing transport remain unchanged.   Some of the literature also includes a long-lived (but not exactly conserved) imbalance mode into the hydrodynamic description \cite{foster, svinstov, narozhny};  for simplicity,  we have not accounted for this effect.   Indeed, the predictions of relativistic hydrodynamics have been confirmed experimentally in \cite{lucas3, crossno};  see also \cite{ghahari, foster2}.

The key advance for the observation of hydrodynamic behavior was the growth of high quality graphene crystals, sandwiched between layers of hexagonal boron nitride.   This dramatically reduced the size and the number of ``charge puddles" -- local inhomogeneity in the chemical potential \cite{xue}.  As a consequence the disorder in graphene became weak enough that hydrodynamic effects were observable at $T\lesssim 100$ K.   (When $T\gtrsim 100$ K,  electron-phonon coupling appears to significantly degrade the electronic energy and momentum,  and hence hydrodynamic behavior).   

Another possibility for limiting the amount of disorder in graphene is to suspend it \cite{bolotin, mayorov}.   The charge puddles in suspended graphene are also inherently quite weak.  However, suspending graphene leads to a new source of disorder:   flexural (out-of-plane) distortions of the graphene crystal.   As we noted in the introduction, these distortions lead to an effective spatial metric $g_{ij}$ given by (\ref{eq:hmetric}).   In the limit where electron-electron interactions are negligible, these flexural modes are known to dominate the resistivity at low temperatures \cite{Mariani}.  Our goal is to understand the implications of these flexural distortions on transport in suspended graphene, in the hydrodynamic limit.   As we have already shown the consequences of (\ref{eq:hmetric}) on transport, our goal here is simply to estimate the size of $h(x,y)$ in suspended graphene, and to comment on whether the hydrodynamic approximation is ever sensible.

In this paper, we will account for the flexural modes by considering motion on a curved space.  In the limit where there are well-defined quasiparticles, it is common to interpret the strain not as the metric deformations  (\ref{eq:hmetric}) but as emergent \emph{magnetic fields} \cite{VozmedianoGraphene,AriasGraphene}.   A priori, this is quite subtle -- a magnetic field breaks time-reversal symmetry, while (\ref{eq:EOM}) preserves time-reversal symmetry.  The resolution to this puzzle is that there are two Dirac points in the Brillouin zone in graphene, and the emergent magnetic field has opposite signs in each valley.   The Dirac fluid of graphene, accounting for both valleys, will remain invariant under time-reversal in the presence of strain.   Nonetheless, as we mentioned previously, strain can open a gap, and so it may be possible that in graphene the presence of strain leads to modifications of the effective hydrodynamics.  These are questions worth considering more carefully in future work.

With these caveats, let us nonetheless estimate the hydrodynamic momentum relaxation rate due to long wavelength flexural fluctuations in graphene.

\subsection{Classical Flexural Phonon Dynamics}

The classical action describing flexural phonons in graphene is \cite{Mariani} \begin{equation}
S = \int \mathrm{d}^2x \mathrm{d}t \left[\frac{\rho}{2}(\partial_t h)^2 - \frac{\kappa}{2}(\partial_i \partial_i h)^2\right] 
\end{equation}
where $h$ is the height of the graphene membrane at position $(x,y)$.   The parameters $\kappa \sim 1$ eV and $\rho \sim 7\times 10^{-7} \; \mathrm{kg/m}^2$ \cite{Mariani2}.     Assuming a square membrane of size $L$, and writing \begin{equation}
h(x,y,t) = \sum_{\mathbf{k}} h_{\mathbf{k}}(t) \mathrm{e}^{\mathrm{i}\mathbf{k}\cdot\mathbf{x}}
\end{equation}  
with allowed wave vectors $\mathbf{k} = 2\mpi/L \times (n_x,n_y)$,  we obtain \begin{equation}
S = \int \mathrm{d}t \sum_{\mathbf{k}} L^2 \left[\frac{\rho}{2} \dot{h}_{\mathbf{k}}\dot{h}_{-\mathbf{k}} - \frac{\rho \omega(\mathbf{k})^2}{2} h_{\mathbf{k}}h_{-\mathbf{k}}\right] 
\end{equation}
As expected, we find a set of decoupled harmonic oscillators with \begin{equation}
\omega(k) \equiv \sqrt{\frac{\kappa}{\rho}} k^2 
\end{equation}

In quantum mechanics, phonons are quantized, and so we should check the length scales at which this classical description will fail.   This occurs when the occupation number of a given phonon mode is comparable to 1, which occurs when $\hbar \omega_{\mathbf{k}}\sim k_{\mathrm{B}}T$.  This occurs when   \begin{equation}
k \gtrsim \sqrt{\frac{k_{\mathrm{B}}T}{\hbar} \sqrt{\frac{\rho}{\kappa}}} \sim \frac{1}{0.2 \; \mathrm{nm}}   \times \sqrt{\frac{T}{100\; \mathrm{K}}}.  \label{eq:kmaxclas}
\end{equation}  
For the remainder of this section, we will restore factors of $\hbar$, $k_{\mathrm{B}}$ etc.   The hydrodynamic description fails when \begin{equation}
k \gtrsim \frac{k_{\mathrm{B}} T}{\hbar v_{\mathrm{F}}} \sim \frac{1}{100 \; \mathrm{nm}} \times \frac{T}{100 \mathrm{K}}  \label{eq:kmaxhydro}
\end{equation}
At any reasonable experimental temperature, there are a classically large number of thermally excited flexural phonons at wave numbers in the hydrodynamic regime.    We also learn that there is a large range of wave numbers where the phonons cannot be treated hydrodynamically.   Hence, we expect a further contribution to momentum relaxation due to these higher wave number phonons, which must be computed using a more microscopic description, such as kinetic theory.    

Next, we must ask whether or not the classical dynamics of flexural phonons is slow enough that the background metric may be treated as static.   The fastest phonon dynamics in the hydrodynamic regime occurs for fluctuations $h_{\mathbf{k}}$ with $\mathbf{k}$ of order (\ref{eq:kmaxhydro}).  Plugging into (\ref{eq:kmaxclas}) we see that the fastest phonon dynamics in the hydrodynamic regime occurs at a rate \begin{equation}
\omega \sim \sqrt{\frac{\kappa}{\rho}} \left( \frac{ k_{\mathrm{B}}T}{\hbar v_{\mathrm{F}}}\right)^2  \sim 10^{-5} \times \frac{T}{100\; \mathrm{K}} \times t_{\mathrm{ee}}.
\end{equation} 
Hence, the metric configuration $h(x,y)$, on hydrodynamic length scales, is essentially frozen in place on electronic time scales, justifying the assumption in our previous hydrodynamic analysis that the background geometry is time-independent.

\subsection{Contribution to Momentum Relaxation Time}
We now compute the contribution of long wavelength fluctuations to the relaxation time for momentum.  First, we must compute the typical size of thermal fluctuations in $h_{\mathbf{k}}$.   Using the classical equipartition theorem, and recalling that $h_{\mathbf{k}}$ contains two independent harmonic oscillators (real and imaginary part): \begin{equation}
\left\langle |h_{\mathbf{k}}|^2 \right\rangle = \frac{2 T}{\kappa k^4 L^2}.   \label{eq:SHOflex}
\end{equation} 
We have once again reverted to natural units.   A straightforward computation, presented in Appendix \ref{appphontime}, reveals that 
\begin{equation}\label{eq:mainphontime}
\frac{1}{\tau} =  \frac{3}{16\mpi^2 } \frac{\eta}{\epsilon+P}  \frac{T^2}{\kappa^2 \xi^2}.
\end{equation}
The hydrodynamic result can only be trusted until for $\xi \lesssim 1/T$, and so we estimate that the contribution of (hydrodynamically) long wavelength flexural phonons to the momentum relaxation time is \begin{equation}
\frac{1}{\tau} \sim \frac{\eta T^4}{\kappa^2(\epsilon+P)}.
\end{equation}
Near the charge neutrality point, the thermodynamic prefactors scale with known powers of temperature, and we obtain $\tau^{-1}\sim T^3$. 

Of course, this must be compared with the other contributions to the momentum relaxation time, including the scattering off of short wavelength phonons.  Using kinetic theory, this has been estimated to  be $\tau^{-1}\sim T^2$ \cite{Mariani2, geimphonon}.  Typically one would account for electron-phonon scattering using kinetic theory,   treating each electron-phonon scattering event as a rare and independent process.     However, we have just seen that in the hydrodynamic limit,  a classical electron fluid with many electron-electron collisions moves in an approximately frozen phonon background.    Thus, one electron can be correlated with the same phonon for over many collisions.  These correlations suggest that the molecular chaos assumption underlying the kinetic description (that scattering events are uncorrelated with each other) could easily break down.

Additional mechanical strain induced by the contacts in a realistic sample of graphene changes the low frequency dispersion relation of flexural modes from quadratic to linear \cite{Mariani2, geimphonon}.   Such a change would alter (\ref{eq:SHOflex}).    But, from the form of (\ref{eq:mainphontime}) it is clear that the smallest wavelength phonons are most efficient at relaxing momentum.  Hence, so long as the quadratic dispersion relation is restored by $k\sim \ell_{\mathrm{ee}}^{-1}$, we expect that (\ref{eq:mainphontime}) approximately accounts for the hydrodynamic contribution to the electron-phonon momentum relaxation rate.

Depending on the nonlinear properties of an elastic membrane,  there can be significant renormalization of the effective $\kappa$ which should be used in (\ref{eq:SHOflex}), due to thermal fluctuations \cite{nelson}.  This effect has been seen recently in molecular dynamics simulations \cite{nelson2} and in experiment \cite{kirigami}.   In a very simple approximation, one estimates that $\kappa_{\mathrm{eff}} \sim \sqrt{TK}/q$ as $q\rightarrow 0$;  $K$ is a constant associated with certain nonlinearities in the elastic free energy.    If this renormalization is significant in the hydrodynamic regime, then we expect that the temperature scaling in $1/\tau$ would be reduced by a factor of approximately $T^3$.

Finally, we note that there are other phonon modes which we could account for.   In particular, there are also in-plane deformations that naturally arise, where the point $x_i$ is displaced to $x_i + d_i(x)$.    In the presence of both a fluctuating height $h(x)$ and displacement $d_i(x)$,  the general expression for $g_{ij}$ is \cite{ciarlet}
\begin{equation}
g_{ij}=\mdelta_{ij}  + \partial_i d_j + \partial_j d_i + \partial_i h \partial_j h.
\end{equation} 
The in-plane phonons of graphene are linearly dispersing, and so $\langle |d_{\mathbf{k}}|^2\rangle \sim k^{-2}$, in contrast to (\ref{eq:SHOflex}).  However, the metric itself depends on $d$, not on $d^2$, and contains one fewer spatial derivative.    Putting this together and generalizing the discussion of Appendix \ref{appphontime}, we estimate that $\tau^{-1} \sim T^4$.   Hence flexural phonons are more important than longitudinal phonons in the hydrodynamic limit.

\section{Conclusion}
In this paper, we have described the effects of inhomogeneous slowly varying strain on hydrodynamic transport in strongly correlated electron fluids.   We have demonstrated that for a (quasi)relativistic system with only strain disorder, the conductivities depend only on the viscosity of the electronic fluid (at least when inhomogeneity is small).   

The conventional theory of electron-phonon scattering estimates the relaxation rate by simply computing low-order Feynman diagrams.  Such an approach is sensible when the mean free path is much larger than the wavelength of both the electrons and the phonons.   However, it is plausible that in charge-neutral graphene and other strongly correlated electron fluids,  the electronic mean free path is short compared to the wavelength of some phonons.   Our hydrodynamic description is the appropriate description of scattering off of these long wavelength phonons,  though we must bear in mind that there will inevitably be a larger number of shorter wavelength phonons, which are not entirely captured by our hydrodynamic model.
%While our formalism accounts for the effect of very long wavelength phonons, including those which are thermally excited,  we do not expect most electron-phonon coupling to be accounted for in our framework, whenever the phonon velocity is small compared to the Fermi velocity. 

A large open problem involves extending the theory of transport beyond the hydrodynamic limit.  In the limit of weak interactions, this can be achieved using kinetic theory:  while challenging, it is possible to completely characterize the ballistic-to-hydrodynamic crossover in this limit \cite{seantocome}.     It would be interesting to understand how the hydrodynamic limit of electron-phonon coupling that we have demonstrated in this work can be understood from such a kinetic theory framework.

\section*{Acknowledgements}
We are grateful to Vadim Cheianov and Philip Kim for useful
discussions.
VS and KS were supported in part
by a VICI (KS) award of the
Netherlands Organization for Scientific Research (NWO), by the
Netherlands Organization for Scientific Research/Ministry of Science
and Education (NWO/OCW), and by
the Foundation for Research into Fundamental Matter (FOM).
AL was supported by the Gordon and Betty Moore Foundation's EPiQS Initiative through Grant GBMF4302.

\begin{appendix}

\section{Onsager Reciprocity}\label{sec:Onsager}
 In this appendix we show that Onsager reciprocity is satisfied on a curved background. This is a non trivial consistency check of our formalism, as it has to be satisfied for any time-reversal symmetric theory of transport. 
 
We begin by introducing some shorthand notation for our proof, following \cite{lucas}.   We denote a uniform spatial average with $\mathbb{E}[X ] = \int \frac{\mathrm{d}^dx}{L^d} \sqrt{g} X$ , where $g$ is the determinant of the spatial metric $g_{ij}$.   Define the vectors \begin{subequations}\begin{align}
F^\alpha_i &\equiv \left(\begin{array}{c} E_i \\ \zeta_i \end{array}\right), \\
\Phi^\alpha &\equiv \left(\begin{array}{c} \mdelta \mu \\ T^{-1}\mdelta T \end{array}\right), \\
\mathcal{J}^\alpha_i &= \left(\begin{array}{c} \mdelta J_i \\ \mdelta Q_i \end{array}\right), \\
\rho^\alpha  &= \left(\begin{array}{c} n \\ Ts \end{array}\right), \\
\Sigma^{\alpha\beta} &= \left(\begin{array}{cc} \sigma_{\textsc{q}} &\  -\sigma_{\textsc{q}}\mu_0 \\ -\sigma_{\textsc{q}}\mu_0 &\ \sigma_{\textsc{q}}\mu_0^2 \end{array}\right), \\
\sigma^{\alpha\beta}_{ij} &=  \left(\begin{array}{cc} \sigma_{ij} &\ T\alpha_{ij} \\ T\bar\alpha_{ij} &\ T\bar\kappa_{ij} \end{array}\right).
\end{align}\end{subequations}
It is straightforward to see that Eqs. (\ref{eq:EOM}) are equal to \begin{subequations}\label{eq:AEOM}\begin{align}
0  &= \nabla_i \mathcal{J}^{\alpha i}  = \nabla_i \left(\rho^\alpha v^i - \Sigma^{\alpha\beta}\nabla_i \Phi^\beta + \Sigma^{\alpha\beta}F^\beta_i\right), \\
0 &= \rho^\alpha \left(\nabla_i \Phi^\alpha  - F^\alpha_i\right) - \nabla^j \left(\eta_{ijkl}\nabla_k v_l\right).
\end{align}\end{subequations}
We have denoted \begin{equation}
\eta_{ijkl} = \eta \left( g_{ik}g_{jl} + g_{il}g_{jk} \right) + \left(\zeta - \frac{2\eta}{d}\right) g_{ij}g_{kl}.
\end{equation} 
To prove Onsager reciprocity we must prove that \begin{equation}
\sigma^{\alpha\beta}_{ij} = \sigma^{\beta\alpha}_{ji}.
\end{equation}
By linearity, we may write the solutions to these equations of motion as 
\begin{subequations}\begin{align}
\Phi^{\alpha} &= \sum_{J=j=1}^d\Phi^{\alpha\beta J} F^{\beta}_j,\\
v^i &= \sum_{J=j=1}^d v^{iJ \beta} F^{\beta}_j.
\end{align}\end{subequations}
We have denoted the index $J$ in capital letters to emphasize that $\Phi^J$ is not a contravariant vector, and that $v^{iJ}$ is a contravariant vector, not a tensor.    (\ref{eq:AEOM}) then becomes  
\begin{subequations}\label{eq:38}\begin{align}
\nabla_i \left(\rho^\alpha v^{iJ \beta} - \Sigma^{\alpha\gamma}\nabla^i \Phi^{\gamma\beta J}\right) &=  -\nabla_i\left(g^{Ji}\Sigma^{\alpha\beta}\right), \\
\rho^\alpha \nabla_i \Phi^{\alpha\beta J} - \nabla^j \left(\eta_{ijkl}\nabla_k v^{\beta J}_l\right) &= \rho^\beta \mdelta^J_i
\end{align}\end{subequations}
Now, by definition \begin{equation}\label{eq:39}
\sigma^{\alpha\beta IJ} = \mathbb{E}\left[\rho^\alpha \mdelta_i^I v^{iJ}_\beta - \Sigma^{\alpha\gamma} g^{Ii}\nabla_i \Phi^{\gamma \beta J}  + \Sigma^{\alpha\beta}\mdelta^{IJ} \right]
\end{equation}
We now use the left hand side of (\ref{eq:38}) to re-write the coefficients of the first two terms of (\ref{eq:39}), after integrating by parts the second term of (\ref{eq:39}) (note that $\mathbb{E}[X\nabla_i Y^i] = -\mathbb{E}[Y^i\nabla_i X]$ -- this can be easily seen using $\nabla_i Y^i = g^{-1/2}\partial_i \left(g^{1/2}Y^i\right)$):  \begin{equation}
\sigma^{\alpha\beta IJ} = \mathbb{E}\left[ \left(\rho^\gamma \nabla_i \Phi^{\gamma\alpha I} - \nabla^n\left(\eta_{inrs}\nabla^r v^s\right) \right) v^{iJ}_\beta + \left(\nabla_k \left(\Sigma^{\gamma\delta}\nabla^k \Phi^{\delta\alpha I} - \rho^\gamma v^{k\alpha I}\right)\right) \Phi^{\gamma \beta J}  + \Sigma^{\alpha\beta}\mdelta^{IJ} \right].
\end{equation}
We next integrate by parts the viscous part of the first term, as well as the second term.   Using the fact that $\mathbb{E}[Y_i \nabla_j Z^{ij}] = -\mathbb{E}[Z^{ij}\nabla_j Y_i]$, this can easily be done.   We find \begin{equation}
\sigma^{\alpha\beta IJ} = \mathbb{E}\left[ \rho^\gamma \nabla_i \Phi^{\gamma\alpha I} v^{iJ \beta}  + \rho^\gamma v^{iI\alpha} \nabla_i \Phi^{\gamma\beta J} + \eta_{inrs} \nabla^r v^{sI\alpha} \nabla^n v^{iJ \beta} -  \Sigma^{\gamma\delta}\nabla^k \Phi^{\delta\alpha I} \nabla_k \Phi^{\gamma \beta J}  + \Sigma^{\alpha\beta}\mdelta^{IJ} \right].
\end{equation}
It is straightforward to determine from this expression, as well as the symmetry properties of $\eta_{ijkl}$,  that $\sigma^{\alpha\beta IJ} = \sigma^{\beta\alpha JI}$, proving Onsager reciprocity.

\section{Numerical Simulations}
\label{app:num}
 The hydrodynamic equations can be solved  on a periodic domain of $L\times L$ grid points with  pseudospectral methods \cite{trefethen}.   The domain is reduced to a set of uniformly distributed grid points $\mathbf{x}_I$ and the continuous partial differential equation \eqref{eq:EOM} is approximated by the discrete form
\begin{equation}\label{eq:pseudo}
\mathsf{L}\mathbf{u}=\mathbf{s}
\end{equation}
where \begin{equation}
\mathbf{u} = \left(\begin{array}{c} \mdelta \mu(\mathbf{x}_I) \\ \mdelta T(\mathbf{x}_I) \\ \mdelta v_i(\mathbf{x}_I) \end{array}\right),
\end{equation}
is the data we wish to solve for,  \begin{equation}
\mathbf{s} =  \left( \begin{array}{c}
-\bar{\nabla}^{i}\sigma_{\textsc{q}}\left(E_{i} - \mu_{0}\zeta_{i}\right)  \\
\bar{\nabla}^{i}\sigma_{\textsc{q}}\mu_{0}\left(E_{i} - \mu_{0}\zeta_{i}\right)  \\
nE_{j}+sT_{0}\zeta_{j}  \end{array} \right),
\end{equation}
is the source for the linearized hydrodynamic equations due to the external fields $E_i$ and $\zeta_i$ (henceforth, the dependence on $\mathbf{x}_I$ of coefficients is implicit),  and  $\mathsf{L}$ is the discretized differential operator \begin{equation}
\mathsf{L} = \left( \begin{array}{ccc}
-\bar{\nabla}^{i}\sigma_{\textsc{q}} \partial_{i} & \bar{\nabla}^{i}\sigma_{\textsc{q}} \frac{\mu_{0}}{T_{0}}\partial_{i} & \bar{\nabla}_i n \\
\bar{\nabla}^{i}\sigma_{\textsc{q}} \mu_{0}\partial_{i} & -\bar{\nabla}^{i} \sigma_{\textsc{q}} \frac{\mu^{2}_{0}}{T_{0}}\partial_{i} & \bar{\nabla}_i s T_{0} \\
n \partial_{j} & s\partial_{i} & -\sqrt{g}^{-1}\partial_k \sqrt{g}\eta(\bar{\nabla}^{k}g_{ji}+\partial_j\delta^k_i) +\eta  (\partial_j g_{kl})\,\bar{\nabla}^{(k}\delta^{l)}_{i} - \partial_{j}(\zeta - \eta)\bar{\nabla}_{i} \end{array} \right).  \label{eq:Lnum}
\end{equation}
The derivative operators in this equation are understood to act on everything to their right, unless contained within parentheses.   They are approximated on the discrete grid using spectral methods.
 By inverting\footnote{The operator $\mathsf{L}$ has two zero eigenvectors, corresponding to a constant shift of $\mu$ and $T$. This can be fixed eliminating an appropriate pair of rows/columns from the matrix $L$, so as to enforce the constraints $\mu(\mathbf{0}) = T(\mathbf{0})=0$.} the system \eqref{eq:pseudo},  we find expressions for $\mdelta \mu$, $\mdelta T$ and $\mdelta v_i$, from which it is simple to compute the spatially averaged charge and heat currents, and hence the thermoelectric conductivity matrix.

The disorder consists of random sine waves both for the chemical potential and for the strain:
\begin{subequations}
\begin{align}
\mu_0(\mathbf{x})&=\bar{\mu_0}+\sum_{|n_x|,|n_y|\leq k} \hat{\mu_0}(n_x,n_y)\sin\left(\phi_x + \frac{2n_x\mpi x }{k\xi}\right)\sin\left(\phi_y + \frac{2n_y\mpi y }{k\xi}\right)\\
h(\mathbf{x})&=\sum_{|n_x|,|n_y|\leq k} \hat{h}(n_x,n_y)\sin\left(\phi'_x + \frac{2n_x\mpi x }{k\xi}\right)\sin\left(\phi'_y + \frac{2n_y\mpi y }{k\xi}\right)
\end{align}\end{subequations}
 where $\bar \mu_0$ is constant, $\phi_{x,y}$ is uniformly distributed on $[0,2\mpi)$ and  $\hat{\mu_0}(n_x,n_y)$, $\hat{h}(n_x,n_y)$ are uniformly distributed on $[-c,c]$, with $c = \sqrt{(2-\mdelta_{n_{x},0}-\mdelta_{n_{y},0})/2}$.   We have chosen this expression for $c$ such that both charge disorder and strain  are not altered by a random homogeneous offset.   $k$ denotes the number of sine waves in a given direction.  In all simulations we have used $L=8k+3$ grid points in the $x$ and $y$ directions.

The numerical simulations were performed using a finite number of disorder modes $k$, both for the charge disorder and for the strain disorder.   This introduces another form of finite size effect to the results.  It can be partially reduced by averaging over multiple disorder realizations.    We also caution that due to the nonlinear way that strain couples to the hydrodynamic equations,  an $h(\mathbf{x})$ which varies on the wavelength $\xi$ leads to a varying metric on the scale $\xi/2$, which can source the equations of motion at even shorter wavelengths due to the nonlinear metric dependence in covariant derivatives.    

\begin{figure}
\pgfplotsset{every axis legend/.append style={at={(.03,0.99)},anchor=north west}}
\pgfplotsset{title style={at={(0.3,0.8)},anchor=north west}}
	\centering
	\includegraphics[scale=0.8]{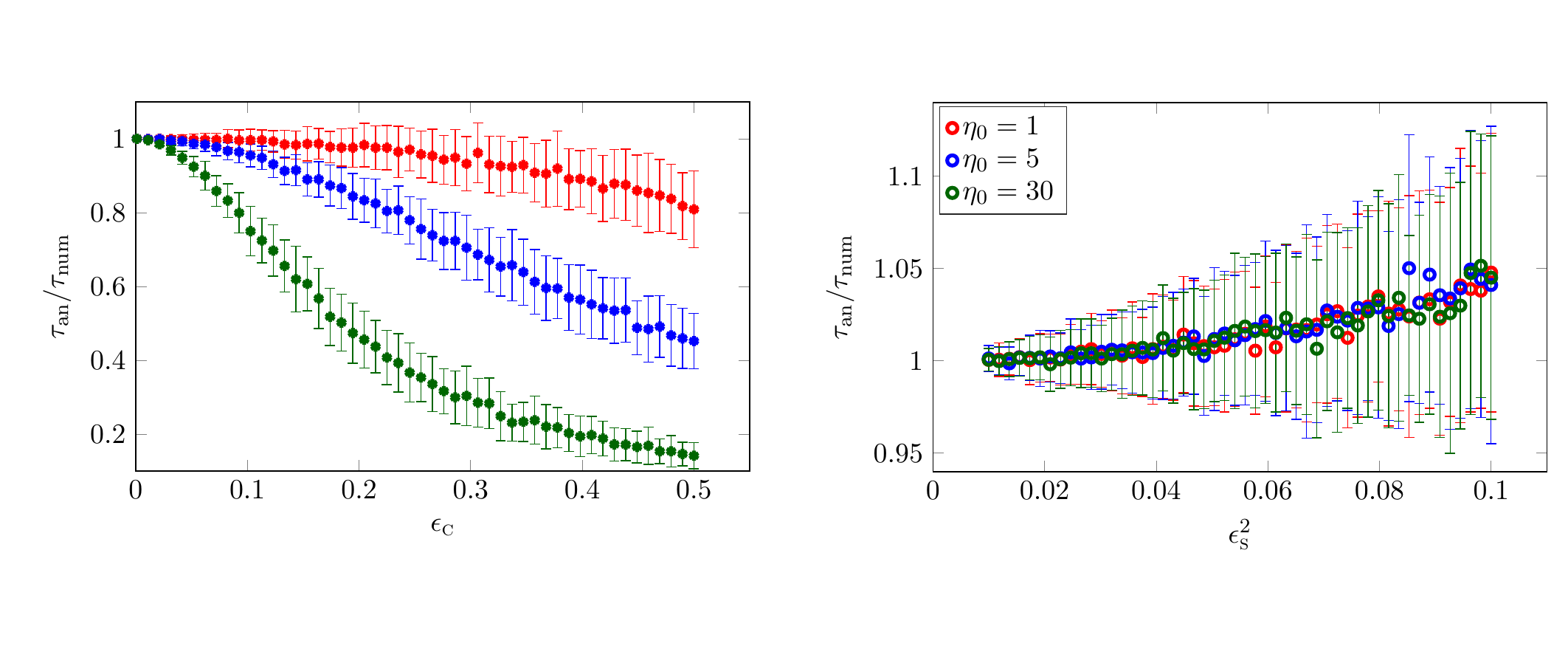}
	\caption{Statistical deviations in the relaxation time for momentum, extracted from the numerically computed conductivity in simulations with $k=2$ and averaged over 100 disorder configurations.   In the left plot, we have assumed all disorder is in the chemical potential;  in the right plot, all disorder is in the metric.   Finite size effects and statistical fluctuations are more significant in the latter case;  this is likely due to the very nonlinear way that strain disorder couples to the hydrodynamic equations.  % Averages have been computed over 100 disorder configurations.
 }
\label{fig:Errorbars}
\end{figure}

Despite these caveats, we expect that our numerical methods converge exponentially quickly upon increasing $L$ \cite{trefethen}. Simple tests confirm this \cite{lucas3}.    We expect therefore that the results presented in the main text are a reasonable quantitative characterization of hydrodynamic transport in the presence of disorder.  We have also found that the deviations in transport coefficients from one realization of disorder to the next is relatively mild:  see Figure \ref{fig:Errorbars} for an example of the size of deviations from one realization of disorder to the next.

%\noindent In order for the code to converge properly, the number of disorder modes for the strain has to be carefully chosen; by inspecting \eqref{eq:NonPertEq}, we observe that the strain $h(\mathbf{x})$ enters in the hydrodynamic equations non linearly, increasing the number of grid points needed. This effect is enhanced with the increasing value of the strain deformation and of its modes. This imply that in the simulations, the grid points number kept  fixed , the increasing of the strain magnitude has to be followed by a decreasing of the strain disorder modes.\\

% The last cause of finite size effect is the finite number of grid points in the pseudospectral method. However we expect an exponential convergence \cite{trefethen} in the number of grid points.\\

\section{Weak Disorder}\label{sec:weakDis}
In this appendix, we perturbatively calculate the transport coefficients assuming that the inhomogeneity is small.  The calculation is analogous to \cite{lucas, lucas3}.    As we discussed in the main text, it is necessary for $\mathrm{\Delta} \mu \sim u$ and $\mathrm{\Delta} h \sim \sqrt{u}$, with $u \rightarrow 0$;  this will lead to transport coefficients which are perturbatively large:  $\sigma_{ij} \sim u^{-2}$.
%In order for the effects of charge and strain disorder to enter the calculation at the same order,  it is necessary that $\mathrm{\Delta} \mu \sim \epsilon_C$ and $h\sim \epsilon_S\sim \sqrt{\epsilon_C}$,  with $\epsilon_C\rightarrow 0$ a perturbatively small parameter, as we will shortly show.    As in \cite{lucas, lucas3}, the response of the fluid is as follows:   at order $\epsilon_C^{-2}$,  there is a constant velocity field $\bar v_i$.    It is this field alone which will determine all transport coefficients.   To compute $\bar v^i$, it is necessary to compute the finite wave number corrections to $\mu$, $T$ and $v^i$  at order $u^{-1}$.   
We will denote with $\hat n(\mathbf{k})$ the charge density at wave number $\mathbf{k}$,  and denote the metric as 
\begin{equation}
g_{ij} = \mdelta_{ij} + \hat g_{ij},
\end{equation}
with $\hat g \sim u$ perturbatively small.   To leading non-trivial order,  $g^{ij} \approx \mdelta^{ij} - \hat g^{ij}$,  with the indices raised trivially using the flat space metric.    

Upon Fourier transforming $\mdelta \mu$, $\mdelta T$ and $\mdelta v_i$, we write \begin{subequations}\begin{align}
\mdelta \mu &= \sum_{\mathbf{k}\ne \mathbf{0}} \mu(\mathbf{k}) \mathrm{e}^{\mathrm{i}\mathbf{k}\cdot\mathbf{x}}, \\ 
\mdelta T &= \sum_{\mathbf{k}\ne \mathbf{0}} T(\mathbf{k}) \mathrm{e}^{\mathrm{i}\mathbf{k}\cdot\mathbf{x}}, \\ 
\mdelta v_i &= \bar v_i +  \sum_{\mathbf{k}\ne \mathbf{0}} v_i(\mathbf{k}) \mathrm{e}^{\mathrm{i}\mathbf{k}\cdot\mathbf{x}}. 
\end{align}\end{subequations}  
We will show that to leading order in $u$,   $\mu(\mathbf{k}) \sim T(\mathbf{k}) \sim v_i(\mathbf{k}) \sim u^{-1}$ while $\bar v_i \sim u^{-2}$.   Let us begin by writing down the $\mathrm{O}(u)$ equations for $\mu(\mathbf{k})$, $T(\mathbf{k})$ and $v^i(\mathbf{k})$.   In these equations, indices may be raised and lowered freely, since all corrections are higher order in $u$.   We obtain \begin{subequations}\label{eq:Sistema}\begin{align}
\mathrm{i}k_{i}n_{0}v^{i}(\mathbf{k})+\sigma_{\textsc{q}0} k^{2}\left(\mu(\mathbf{k}) - \frac{\mu_{0}}{T_{0}}T(\mathbf{k})\right) + \mathrm{i} k_{i}\bar{v}^{i}\hat{n}(\mathbf{k})
+  \mathrm{i} k_{i}\bar{v}^{i}n_{0}\frac{\hat{g}_{kk}(\mathbf{k})}{2}&=0,\\
\mathrm{i}k_{i}T_{0}s_{0}v^{i}(\mathbf{k})-\sigma_{\textsc{q}0} k^{2}\mu_{0}\left(\mu(\mathbf{k}) - \frac{\mu_{0}}{T_{0}}T(\mathbf{k})\right) + \mathrm{i} k_{i}\bar{v}^{i}\hat{s}(\mathbf{k})T_{0}
+  \mathrm{i} k_{i}\bar{v}^{i}s_{0}T_{0}\frac{\hat{g}_{kk}(\mathbf{k})}{2}&=0,\\
\mathrm{i}k_{i}n_{0}\mu(\mathbf{k})+\mathrm{i}k_{i}s_{0}T(\mathbf{k})+\eta_0 k^2v^{i}(\mathbf{k})+\zeta_0 k_{i}k_{j}v^{j}(\mathbf{k})
+\bar{v}^{j}k_{j}\left[\eta_0 k_{k}\hat{g}_{ki}(\mathbf{k})+(\zeta_0-\eta_0)k_{i}\frac{\hat{g}_{kk}}{2}\right]&=0\label{eq:momentum}
\end{align}\end{subequations}
Upon solving for $T(\mathbf{k})$, $\mu(\mathbf{k})$ and $k_i v_i(\mathbf{k})$ at leading order, we obtain: 
\begin{subequations}\label{eq:perturbSol}\begin{align}
T(\mathbf{k})&=-\frac{\mathrm{i}k_{i}\bar{v}_{i}}{\sigma_{\textsc{q}0}k^{2}(\epsilon_{0}+P_{0})^{2}}\left(\sigma_{\textsc{q}0}k^{2}(\eta_{0}+\zeta_{0})(\mu_{0}\hat{n}(\mathbf{k})+T_{0}\hat{s}(\mathbf{k}))T_{0}-T_{0}n_{0}(T_{0}s_{0}\hat{n}(\mathbf{k})-T_{0}n_{0}\hat{s}(\mathbf{k}))\right)+ \notag \\
&+\mathrm{i}T_0 \frac{k_i \bar{v}_i}{k^2 (\epsilon_0 + P_0)}\eta_0\left[ k_j \hat{g}_{jk}(\mathbf{k})k_k  - k^2 \hat{g}_{ll}(\mathbf{k})\right],\label{eq:T(k)}\\
\mu(\mathbf{k})&=-\frac{\mathrm{i}k_{i}\bar{v}_{i}}{\sigma_{\textsc{q}0}k^{2}(\epsilon_{0}+P_{0})^{2}}\left(\sigma_{\textsc{q}0}k^{2}(\eta_{0}+\zeta_{0})(\mu_{0}\hat{n}(\mathbf{k})+T_{0}\hat{s}(\mathbf{k}))\mu_{0}+T_{0}s_{0}(T_{0}s_{0}\hat{n}(\mathbf{k})-T_{0}n_{0}\hat{s}(\mathbf{k}))\right)+\notag  \\
&+\mathrm{i}\mu_0 \frac{k_i \bar{v}_i}{k^2 (\epsilon_0 + P_0)}\eta_0\left[ k_j \hat{g}_{jk}(\mathbf{k})k_k  - k^2 \hat{g}_{ll}(\mathbf{k})\right],\label{eq:mu(k)}\\
k_{i}v^{i}(\mathbf{k})&=-\frac{k_{i}\bar{v}^{i}}{\epsilon_{0}+P_{0}}\left(\mu_{0}\hat{n}(\mathbf{k}) + T_{0}\hat{s}(\mathbf{k})\right)- k_i \bar{v}^i \frac{\hat{g}_{kk}(\mathbf{k})}{2}.\label{eq:v(k)}
\end{align}
\end{subequations}
It is also helpful to combine these equations to obtain \begin{equation}
\eta_0 k^2 v_i(\mathbf{k}) = \eta_0 \left(k_i \frac{\hat g_{ll}}{2} - k_k \hat g_{ki} - k_i \mathbb{P}_{mn}\hat g_{mn} - \frac{\mu_0\hat n  +T_0 \hat s}{\epsilon_0+P_0}k_i\right) k_j \bar v_j .   \label{eq:etavi}
\end{equation}

We now spatially average the momentum conservation equation at $\mathrm{O}(u^0)$, and find \begin{align}
n_0 E_i + T_0 s_0 \zeta_i &= \sum_{\mathbf{k}} \left[\hat n(-\mathbf{k}) \mathrm{i}k_i \mu(\mathbf{k}) + \hat s(-\mathbf{k}) \mathrm{i}k_i T(\mathbf{k}) \right. \notag \\
&\;\;\;\; \left.+ \frac{\eta_0}{2} \mathrm{i}k_k \hat g_{ll}(-\mathbf{k}) \left((\nabla^k v_i)(\mathbf{k}) + (\nabla^i v_k)(\mathbf{k})\right)- \mathrm{i}k_i  \eta_0 \hat g_{kl}(-\mathbf{k}) (\nabla^k v^l)(\mathbf{k}) \right]
\end{align}
It is helpful to spatially average the last column of (\ref{eq:Lnum}), without an overall metric factor, to obtain this equation, as many total derivative terms are shown explicitly.  Next, we note that in all terms in this equation, we may raise and lower spatial indices with the flat space metric, so long as we keep in mind \begin{equation}
(\nabla_k v^i)(\mathbf{k})  = \mathrm{i}k_k v^i + \frac{\mathrm{i}}{2}\left(k_j \hat g_{ik} + k_k \hat g_{ij} - k_i \hat g_{jk}\right) \bar v_j + \cdots. 
\end{equation}
Our goal is to eliminate $\mu(\mathbf{k})$, $T(\mathbf{k})$ and $v_i(\mathbf{k})$ in favor of $\bar v_j$.   Given (\ref{eq:perturbSol}) and (\ref{eq:etavi}) this is straightforward but tedious.   One finds that \begin{equation}
n_0E_i + T_0s_0\zeta_i = (\epsilon_0+P_0) \left[\left( \tau_{ij}^{-1}\right)^{(\mu\mu)} +  \left(\tau_{ij}^{-1}\right)^{(\mu h)} +  \left(\tau_{ij}^{-1}\right)^{(hh)}\right]\bar v_j  \label{eq:55}
\end{equation}
where \begin{subequations}\begin{align}
\left( \tau_{ij}^{-1}\right)^{(\mu\mu)} &= \sum_{\mathbf{k}} \frac{k_i k_j}{k^2} \frac{\left|T_0 n_0 \hat s(\mathbf{k}) - T_0 s_0 \hat n(\mathbf{k})\right|^2 + k^2 \sigma_{\textsc{q}}(\eta_0+\zeta_0) \left|T_0 \hat s(\mathbf{k}) + \mu_0 \hat n(\mathbf{k})\right|^2}{\sigma_{\textsc{q}}(\epsilon_0 + P_0)^3}  \\
\left( \tau_{ij}^{-1}\right)^{(\mu h)} &= 2\eta_0  \sum_{\mathbf{k}}  k_i k_j \, \frac{\bar\mu_0 n(\mathbf{k})+T_0 s(\mathbf{k})}{(\epsilon_0 + P_0)^2} \hat{g}_{kl}(-\mathbf{k})\, \mathbb{P}_{kl},  \\
\left( \tau_{ij}^{-1}\right)^{(hh)}  &= \frac{\eta_0}{2(\epsilon_0+P_0)} k_i k_j \left[\left|\hat g_{ll}(\mathbf{k})\right|^2 + \left|\hat g_{kl}(\mathbf{k})\right|^2 + 2\left|\frac{k_kk_l}{k^2}\hat g_{kl}(\mathbf{k})\right|^2 \right. \notag \\
&\;\;\;\; \left. -2 \hat g_{mm}(-\mathbf{k}) \frac{k_k k_l}{k^2}\hat g_{kl}(\mathbf{k}) - 2\frac{k_mk_n}{k^2}\hat g_{ml}(-\mathbf{k}) g_{nl}(\mathbf{k})\right] .
\end{align}\end{subequations}
From (\ref{eq:55}) we obtain an expression for $\bar v_j$.  Using $J_i \approx n_0 \bar v_i$ and $Q_i \approx T_0 s_0 \bar v_i$, we obtain (\ref{eq:TranspMain}) and (\ref{eq:scatteringRate}).

\subsection{Phonon Contribution} \label{appphontime}
Let us now evaluate the contribution of flexural phonons to the momentum relaxation time.
In order to evaluate the momentum relaxation time, we need to evaluate thermal averages which can be written in the generic form \begin{align}
&\sum_{\mathbf{q},\mathbf{r}} \left\langle k_i k_j q_k h_{\mathbf{q}} (q-k)_l h_{\mathbf{k}-\mathbf{q}} r_m h_{\mathbf{r}} (r+k)_n h_{-\mathbf{r}-\mathbf{k}} \right\rangle \times \frac{1}{2}\left(\mathbb{P}_{mn}\mathbb{P}_{kl} + \mathbb{P}_{km}\mathbb{P}_{ln}\right) \notag \\
 &= 2\sum_{\mathbf{q}} k_i k_j q_k q_m (q-k)_l (q-k)_n \left(\frac{2T}{\kappa L^2}\right)^2 \frac{1}{(q^2(q-k)^2)^2}\times \frac{1}{2}\left(\mathbb{P}_{mn}\mathbb{P}_{kl} + \mathbb{P}_{km}\mathbb{P}_{ln}\right) \notag \\
  &= 2\sum_{\mathbf{q}} k_i k_j q_k q_m q_l q_n \left(\frac{2T}{\kappa L^2}\right)^2 \frac{1}{(q^2(q-k)^2)^2} \times \frac{1}{2}\left(\mathbb{P}_{mn}\mathbb{P}_{kl} + \mathbb{P}_{km}\mathbb{P}_{ln}\right) .
\end{align}
In the second line, we have used (\ref{eq:SHOflex}) to remove thermal averages of $h_{\mathbf{k}}$, and in the third line we have used the fact that $k_i \mathbb{P}_{ij} = 0$.   Now, we need to evaluate the sum over $\mathbf{q}$.   In general, we will find
\begin{equation}
\sum_{\mathbf{q}} \frac{q_k q_l q_m q_n}{(q^2(q-k)^2)^2} = A\left(\mdelta_{ij}\mdelta_{kl} + \mdelta_{ik} \mdelta_{jl} + \mdelta_{il}\mdelta_{jk}\right) + B\left(\frac{k_ik_j}{k^2}\mdelta_{kl} + \text{5 permutations}\right) + C\frac{k_ik_jk_kk_l}{(k^2)^2}  \label{eq:sumqgen}
\end{equation} 
but due to the projectors coming from the formula for $\tau^{-1}$, we need only compute the coefficient $A$.    We do so by assuming that $\mathbf{k} = k\hat{\mathbf{x}}$ and then set all indices $k=l=m=n=y$ in (\ref{eq:sumqgen}).    In order to approximate the sum over $\mathbf{q}$ with an integral, we employ the standard substitution \begin{equation}
\sum_{\mathbf{q}} \rightarrow  L^2 \int \frac{\mathrm{d}^2\mathbf{q}}{4\mpi^2}.
\end{equation}
Hence, we find \begin{align}
\sum_{\mathbf{q}} \frac{q_y^4}{(q^2(q-k)^2)^2} &= 3A \approx \frac{L^2}{4\mpi^2} \int\limits_0^\infty  \mathrm{d}q \int\limits_0^{2\mpi} \mathrm{d}\theta \; \frac{q\sin^4\theta}{\left(q^2+k^2-2kq\cos\theta\right)^2} \notag \\
&= \frac{L^2}{4\mpi^2} \int\limits_0^\infty  \mathrm{d}q  \frac{3\mpi q}{8k^4q^4}\left(q^4+k^4 - \left|k^4-q^4\right|\right) = 3\frac{L^2}{4\mpi^2} \times \frac{\mpi}{4k^2}.
\end{align}
Hence, we find a finite value for $A$.   In contrast to $A$,  one can show that other coefficients in (\ref{eq:sumqgen}) do have logarithmic divergences.    Given the value for $A$, we find \begin{align}
&2\sum_{\mathbf{q}} k_i k_j q_k q_m q_l q_n \left(\frac{2T}{\kappa L^2}\right)^2 \frac{1}{(q^2(q-k)^2)^2} \times \frac{1}{2}\left(\mathbb{P}_{mn}\mathbb{P}_{kl} + \mathbb{P}_{km}\mathbb{P}_{ln}\right) \notag \\
&= k_i k_j  \left(\frac{2T}{\kappa L^2}\right)^2\left(\mathbb{P}_{mn}\mathbb{P}_{kl} + \mathbb{P}_{km}\mathbb{P}_{ln}\right)\left(\mdelta_{ij}\mdelta_{kl} + \mdelta_{ik} \mdelta_{jl} + \mdelta_{il}\mdelta_{jk}\right) \frac{L^2}{16\mpi k^2}  = \frac{1}{L^2} \frac{3 k_ik_j}{2\mpi k^2}  \frac{T^2}{\kappa^2}
\end{align}
Employing \begin{equation}
\sum_{\mathbf{k}} \frac{k_ik_j}{k^2} \approx \frac{L^2}{8\mpi \xi^2} \mdelta_{ij},
\end{equation}
we find \eqref{eq:mainphontime}.

\end{appendix}
\bibliographystyle{unsrt}
\addcontentsline{toc}{section}{References}
\bibliography{theorypaperbib}

\end{document}